\RequirePackage{fix-cm}
\documentclass[natbib]{svjour3}                     %
\smartqed  
\usepackage{graphicx}
\usepackage{epsf}
\usepackage{amssymb}
\usepackage{amsmath}
\usepackage{placeins}
\usepackage{color}
\usepackage{hyperref}
 \journalname{SSRv}
\newcommand{\be}{\begin{equation}}
\newcommand{\ee}{\end{equation}}
\newcommand{\beq}{\begin{eqnarray}}
\newcommand{\eeq}{\end{eqnarray}}
\newcommand\subsun[1]{{$_{\normalsize\odot}$}}

\begin{document}
\title{Circumstellar interaction in supernovae in dense environments\\ - an observational perspective}

\titlerunning{Circumstellar interaction in supernovae}        

\author{Poonam~Chandra$^{1,2}$ }

\institute{
$^1$ National Centre for Radio Astrophysics, Tata Institute of Fundamental Research, Pune University Campus, Ganeshkhind, Pune 411 007, INDIA\\
$^2$ Department of Astronomy, Stockholm University, AlbaNova, SE-106 91 Stockholm,
SWEDEN\\
\email{poonam@ncra.tifr.res.in}\\
             }

\date{Received: date / Accepted: date}

\maketitle

\abstract

In a supernova explosion, the ejecta interacting with the surrounding circumstellar medium (CSM) give rise to variety of radiation. Since CSM is created from the mass lost from the progenitor star, it carries footprints of the late time evolution of the star. This is one of the unique ways to get a handle on the nature of the progenitor star system. Here, I will focus mainly on the supernovae (SNe)  exploding in dense environments, a.k.a. Type IIn SNe. 
 Radio and X-ray emission from this class of SNe have revealed important modifications in their radiation properties, due to the presence of high density CSM.  Forward shock dominance of the X-ray emission, internal free-free absorption of the radio emission, episodic or non-steady mass loss rate, and asymmetry in the explosion seem to be common properties of this class of SNe.  


\section{Introduction}
\label{introduction}

A massive star (mass $M >8 \rm M_\odot$)  evolves for millions of years, burning nuclear fuel, and keeping the star stable with resulting outward radiation pressure.
 However, when the  nuclear fuel in the star is exhausted, the equilibrium between
gravity and radiation pressure ceases to exist  and the star collapses 
under its own gravity within a fraction of a
second.  At some point  the implosion turns into an explosion, and a core-collapse supernova (SN) is born, leaving behind
a relic which is either a neutron star or a black hole.  This is the most simplistic picture  of a SN explosion.  The classic paper by  \citet{bb85}  remains one of the best resources to get  overall 
qualitative picture of the SN explosion.
Understanding the detailed and quantitative characteristics
 of the explosion  requires  complex neutrino physics, general relativity and magnetohydrodynamics.
 Other kinds of supernovae (SNe), thermonuclear SNe, arising from the detonation of  a white dwarf  residing in a binary system are not the subject of this 
 review.

Stars lose mass from their least gravitationally bound outermost layers.  The mass loss rate  $\dot M$ from a massive star can be expressed as
 $\dot M  = 4\pi r_\infty^2 \rho(r_\infty) v_{\infty}$,  where $\rho(r_\infty)$ is the average mass density  of the star at a distance $r_\infty$ from the centre where
the wind speed $v_{\rm wind}$ has reached terminal velocity $v_{\rm \infty}$ \citep{smith14b}. 
While the Sun is losing mass at a rate of $\dot M \sim 10^{-14} \,\rm M_\odot\,\rm yr^{-1}$,  the massive stars
lose mass more profusely owing to  their much larger radii. 
The mass lost manifests 
itself in the form of dense winds moving with velocities of the order 10--1000 km s$^{-1}$, creating a high density medium surrounding the star, a.k.a.  circumstellar medium (CSM).
The wind  speed is generally  proportional to the star's escape velocity, thus the outflows from the  yellow super giants (YSGs) and red super giants (RSGs)  with larger radii are normally slower
($\sim 10-20$ km s$^{-1}$), whereas luminous blue variable (LBV)  and blue super giants (BSGs)  with  smaller radii have faster winds
 ($\sim 100$ km s$^{-1}$).
In reality, though,  high radiation from the  SN can potentially accelerate the CSM winds to much higher speeds \citep{smith14b}.  

In normal core-collapse SNe, the spectra and energetics are mainly governed by the explosion dynamics, and do not have imprints from the CSM.
However, in a class of core-collapse SNe exploding in dense environments, termed as type IIn SNe  (hereafter SNe IIn),  the   high CSM densities reveal  their presence in
the  form of narrow emission lines in their optical spectra.  
While the radioactive decay mainly powers the optical light curves in normal core collapse SNe,  the  energetics have a significant
contribution from the SN explosion ejecta interacting with the dense CSM in SNe IIn, resulting in high bolometric and H-$\alpha$ luminosities  \citep{chugai90}. 
Due to this extra source of energy, they are detectable up to cosmological distances. 
The farthest detected Type IIn SN  is at a 
redshift, $z=2.36$ \citep{cooke+09}. SNe IIn are the focus of this review.

\section{Progenitors of SNe IIn}
\label{progenitors}

The progenitor  mass loss rates required to explain the extreme densities of SNe IIn are 
quite high, e.g. $10^{-3}-10^{-1}  \rm M_\odot\,\rm yr^{-1}$.
 \citet{chu04} derived mass loss rate for SN 1994W to be $\sim 0.2\, M_\odot\,{\rm yr^{-1}}$, whereas, the mass loss
rate for SN 1995G was found  to  be $\sim 0.1\, M_\odot\,{\rm yr^{-1}}$  \citep{chu03}.  In case of SN 1997ab, \citet{sal98} argued that 
 narrow  P-Cygni profile  superimposed on the broad emission lines implied a mass loss rate of $\sim 10^{-2}\, M_\odot\,{\rm yr^{-1}}$ for a presupernova
wind  velocity of $90\,{\rm km\, s^{-1}}$. \citet{chandra+12b} and \citet{chandra+15} found  mass loss rates $ \sim 10^{-3} \rm M_\odot\,\rm yr^{-1}$ and 
 $ \sim 10^{-1} \rm M_\odot\,\rm yr^{-1}$  for SN 2006jd and SN 2010jl, respectively.
This indicates that the mass loss rates in SNe IIn are   extremely high and  not explained in standard stellar evolution theories of normal class of evolved stars,
 such as Wolf-Rayet (WR)
 stars, LBVs, YSGs, RSGs etc \citep{fullerton+06}.  One possibility is that such high mass loss rates may be 
related to
an explosive event a few years before the SN
outburst \citep{chu03,pas07}. 
Radiation production close to the Eddington limit can  make the star unstable, which can lead to eruptive  mass loss in a dramatically  short time. However,  the only stars which could 
  incorporate enhanced mass loss shortly before explosion are 
super-asymptotic giant branch stars (AGB; mass $8-10\,M_\odot$), massive RSGs (mass $17-25\,M_\odot$) and LBV giant eruptions (mass $>35 M_\odot$) \citep{fullerton+06}.

While erupting LBV progenitors are favorite  models for bright SNe IIn because of their extreme mass loss rates \citep{humphreys+99}, they are a transient phase between O-type and WR stars \citep{humphreysdavidson94}, and
 are not  supposed to explode at this phase. Even if they explode, the fine tuning of the explosion at
the core coinciding with the  enhanced episodic mass loss from the outer most layer  is not understood. 
Yet  archival data in some SNe IIn seem to  suggest LBVs as progenitors,  in  particular    SN 2005gl \citep{galyam+07}, SN 2009ip \citep{mauerhan+13a}, 
and  SN 2010jl \citep{smith+11}.  The pre-explosion Hubble Space Telescope (HST) observations suggested an LBV progenitor for SN 2005gl \citep{galyam+07}.  In addition,
 the inferred pre-shock wind speed of  $420$ km s$^{-1}$ deduced from narrow hydrogen lines was also found to be consistent with the speeds of  LBV
eruptive  winds \citep{smith+10}.
In SN 2009ip, multiple pre-SN eruptions were  observed before the star was  finally thought to explode as SN IIn in 2012, strengthening the connection between
SNe IIn and   an LBV  progenitors
\citep{mauerhan+13a}. However,  whether the 2012 event was a true SN is still debated \citep{fraser13}. 
In case of SN 2010jl,  pre-explosion archival HST data   revealed a luminous, blue point source at the position of the SN, suggesting $M>30M_\odot$ LBV progenitor \citep{smith+11}. 
But recently \cite{fox+17} presented HST Wide-Field  Camera (WFC3) imaging of the SN 2010jl field obtained $4-6$ years after the SN explosion and showed that the SN is 0.61'' 
offset from the blue source previously considered   the progenitor LBV star.
The pre-SN eruptions  have  also  been looked for and were found in roughly half the SNe, in a sample of $\sim 16$ SNe IIn from  the Palomar Transient Factory  archival data
\citep{ofek+14}.  However, these arguments were contradicted by
 \citet{bilinsky+15}, who analysed  12 years of Katzman Automatic Imaging Telescope archival data having five known SNe IIn locations and found no precursors.
 \citet{anderson+12} used H-$\alpha$ emission and near-ultraviolet (UV) emission as tracers of ongoing ($<16$ Myr old)  and recent ($16-100$ Myr old) star formation,
 respectively.   He used pixel statistics to build distributions of associations of different types of SNe  with host galaxy star formation. His sample of 19 SNe IIn was found
to  trace the recent star formation but not the ongoing star formation. Under the assumption that more massive a stars have  shorter  life times,  this implied that
contrary to  general believe, the  majority of SNe IIn do not arise from the most massive stars.  This was also suggested by \citet{cd94} based on the 
low ejecta mass $M_{\rm ej} <1 M_\odot $ in SN 1988Z. In addition, \citet{sanders+13}   reported the detection of another interactive kind (Type Ibn) of SN,
  PS1-12sk,  in a giant elliptical galaxy CGCG 208-042
 ($z = 0.054$),
with no evidence for ongoing star formation at the explosion site.  This further challenges the idea of interacting SNe arising
from  very massive progenitor stars, and puts a question mark on the LBV progenitors.

  In addition, questions have been raised over the true nature of LBVs. 
\cite{smith15a} challenged the current understanding of LBVs, i.e. they  being a transient phase between O-type and WR stars. They found that the LBVs are not spatially located in O-type clusters and are statistically more isolated than O-type 
and WR stars. Based on the relatively isolated environments of LBVs, they 
 suggested that LBVs are the  product of binary evolution. The explained LBVs to be evolved massive blue stragglers. 
 \citet{smith16b} took late time HST observations of the SN 2009ip site to search for evidence of recent star formation in the local environment, and found none within a
 few kpc. If LBVs are
indeed a transient phase between O-type and WR stars, they should be found in the massive star forming regions.  
They proposed that  the SN 2009ip  progenitor may have been  the product of a merger or binary mass transfer, rejuvenated after  4--5 Myr.

Even though LBVs  remain the most favoured and exotic progenitors of SNe IIn, alternative models too have been suggested.  \citet{cd94,chevalier12} proposed a binary model where common envelope evolution can give rise to SNe IIn. In this model, the SN is triggered by the inspiral of a compact neutron star or black hole to the central core of the binary star. Pulsational pair instability models too have been suggested, especially in Type IIn SLSNe, e.g.  SN 2006gy
\citep{woosley07}.  \citet{qs12} proposed a graveity-wave driven mass loss model for such SNe IIn. According to this model, the fusion luminosity post C-burning stage can reach super-Eddington values and generate  gravity wave in massive stars.  As the gravity waves propagate towards stellar surface, they can convert into sound waves and dissipation of the sound waves can
trigger a sudden enhanced mass loss.

Overall, there is a large discrepancy in the progenitor models of SNe IIn. The heart of the problem is that  SNe IIn classification is an external one, where any  
explosion surrounded by
very high density can mimic a type IIn SN.   Thus SNe IIn are governed by external factors rather than by internal explosion dynamics. 
Recently some superluminous supernovae (SLSNe) have also been explained
in the  extreme ejecta-CS interaction scenario by \cite{ci11}. 
These are expected to be the extreme  version of  SNe IIn \citep{quimby+13}.
Thus  SNe IIn certainly encompass
objects of different stellar evolution and mass loss history.
For example, SN 1994W had a very fast decline after  about $> 125$ days  \citep{sollerman+98}, contrary to  SN 1988Z
which remained bright for a very long time  \citep{turatto+93}. 
SN 1998S was initially classified as a  bright Type IIL SN, 
however, at a very early stage ($<$ 10 days) it started to show characteristics 
of type  IIn SN  \citep{li+98}.
 A special class of SNe IIn is characterized
by spectral similarities to Type Ia SNe at
peak light, but later shows SN IIn properties, e.g., SNe 2002ic,  2005gj, PTF 11kx. While the  explosion mechanism is not determined in the first two SNe,  PTF 11kx was clearly
considered to be a thermonuclear event \citep{dilday+12}.
Some SNe IIn also appear to be related to Type Ib/c SNe.  
For example,  SNe 2001em and 2014C were initially classified as type Ib SNe, however, at late epochs, signatures of the  dense CSM interaction were found and these SNe were reclassified as SNe IIn \citep{cc06,milisavljevic+15}.
These are not the only SNe Ib/c to show late time CSM interaction signatures. While analysing a sample of 183 SNe Ib/c,  \citet{margutti+17} found  these late rebrightenings
in SNe  2003gk, 2007bg, and PTF11qcj as well.

A well sampled long term observational follow up of ejecta-CSM interaction is the key to unravel the true nature of the SNe IIn progenitors through their  footprints in their
light curve and spectra.

\section{Circumstellar Interaction}
\label{CSM}

 After a SN explosion, the shock wave propagates through the star and after shock breakout the expanding ejecta start to interact with the surrounding medium. 
 While the inner part of the ejecta density profile
is flatter, it is less important for the ejecta-CSM interaction dynamics. The outer ejecta density profile,  more relevant part for the interaction,  can be described as 
$ \rho_{\rm ej} = \rho_o (t/t_o)^{-3} (V_ot/r)^n$ \citep[where for a mass element $m_i$, ejecta velocity $V(m_i)$, radius $r(m_i)$ and 
density $\rho(m_i)$ have time dependence of  $r(m_i)=V(m_i)t$ and $\rho(m_i) = \rho_o(m_i)(t_o/t)^3$,][]{cf16}.
The factor  $t^{-3}$ comes due to free expansion of the gas.

The ejecta shock moving through stellar layers is radiation dominated mediated by photons.
However, 
eventually 
the photons will  diffuse out, leading to the disappearance of the radiation dominated shock and acceleration of the outer gas. The velocity of the accelerated outer gas  drops with radius because of 
the decreasing flux, hence the inner gas would catch up with the outer gas. Since the ejecta velocity is still supersonic, a viscous shock will be formed, which will commence the interaction with the surrounding medium.

Diffusion of the photons will happen when the optical depth is smaller than the ratio of speed of light, $c$ to ejecta velocity $V_{\rm ej}$, i.e. $\le c/V_{\rm ej}$ and 
the photon diffusion time scale  
becomes comparable to that of age of the SN \citep{cf16}.
In normal core-collapse SNe, the  diffusion happens at the progenitor radius,  in the form of   shock breakout and a huge amount of energy \citep[$10^{46}-10^{48}$ ergs,][]{mm99,cf16} is released, mainly in extreme ultraviolet (EUV) to soft X-rays. 
The shock breakout time scale is small ($\sim 1$ hr), and is followed by the optical peak due to radioactivity. 
However, in some SNe IIn (especially superluminous SNe IIn), the density  of the medium surrounding the exploding star is so high that
optical depth $\tau_{\rm wind} > c/V_{\rm ej}$, and the shock breakout will happen
 in the CSM over a longer period of time, powering the bolometric  
light curve.  Radiation from shock breakout can ionize and accelerate the immediate medium around it. 
\citet{svirski+12} have predicted the generation of late time hard X-ray emission in these prolongated shock breakouts, which has been detected in a few SNe 
\citep{ofek+13}.

 \begin{figure}
\begin{center}  
 \includegraphics[angle=0,width=0.98\textwidth]{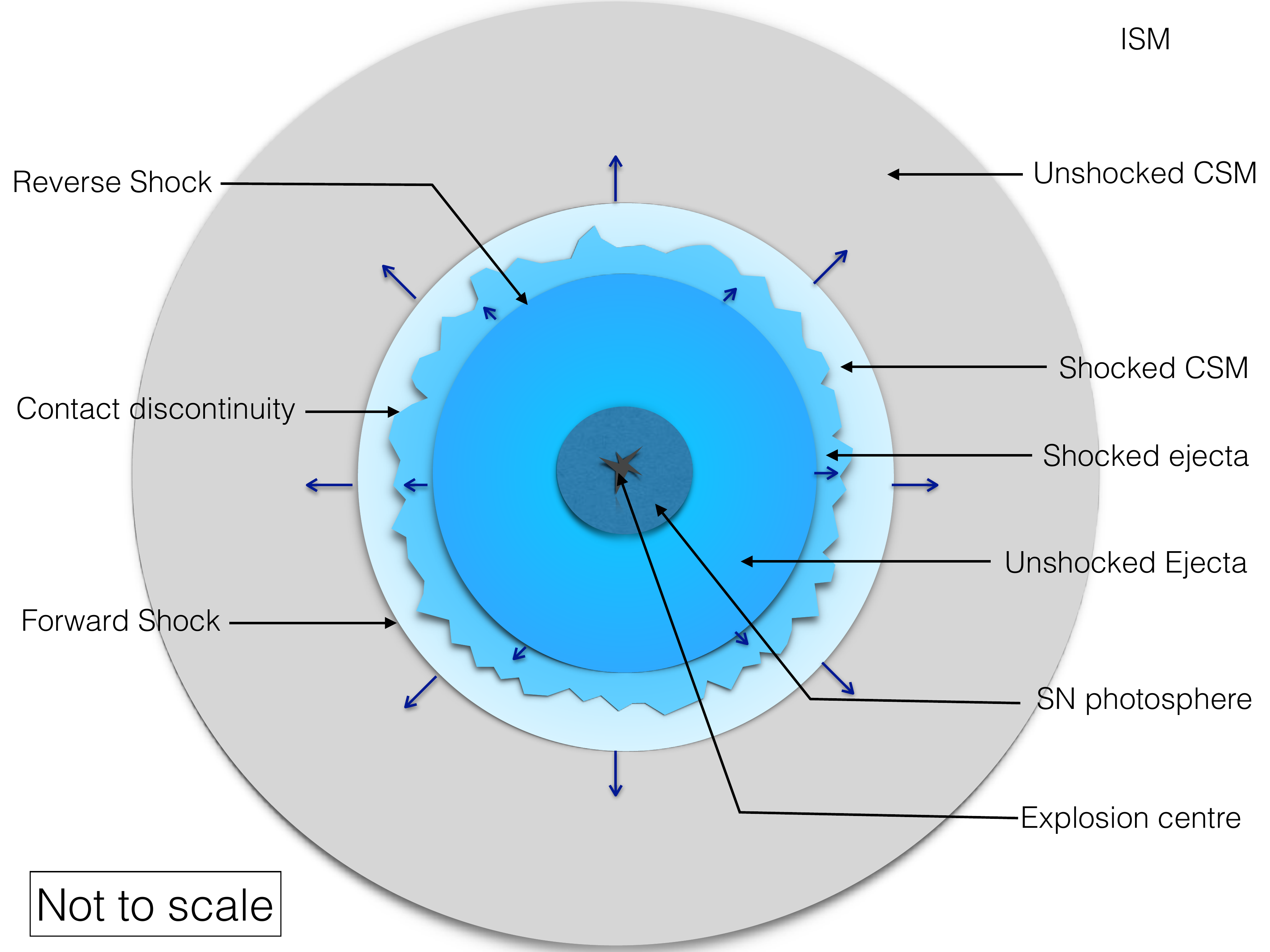}
 \caption{Schematic diagram of a SN ejecta interacting with the surrounding CSM. This creates a hot $10^9$~K forward shock, a $10^7$~K reverse shock and a 
 contact discontinuity in between. Hot forward and reverse shocks produce X-ray emission, whereas the electrons accelerated in the forward shock in the presence of
 the enhanced magnetic field produce synchrotron radio emission. Reprocessed X-ray can also come out as optical and UV radiation.}
   \label{fig:csm}
\end{center}
\end{figure} 

The structure of the unshocked CSM is important for the further ejecta-CSM interaction post shock-breakout. 
The CSM density profile $\rho_{\rm wind}$
 depends upon mass loss rate  as   $\rho_{\rm wind}(r)=\dot M/4\pi r^{-s} v_{\rm wind}$ (where $s=2$ for a steady wind). Thus if one has information of the 
$\rho_{\rm wind}$,  the mass loss rate of the progenitor star can be estimated, provided the wind velocity is known. 
The  fast moving ejecta  collides with the 
 CSM leading to the formation of a   `forward' 
shock moving in the CSM  and a `reverse'
shock propagating back into
the stellar envelope {\it relative} to
the {\it expanding}
stellar ejecta \citep[Fig. \ref{fig:csm} and ][]{chevalier81,chevalier82a}.  In between the forward and the reverse shocks there exists a contact discontinuity, which is anything 
but smooth due to the  instabilities in the region caused by low density shocked CSM decelerating high density
shocked ejecta.
The forward shock has a velocity of the order $10,000$ km s$^{-1}$ and  temperature $ \sim10^9$ K, whereas the reverse shock moves 
 with  velocity $\sim$ 1000 km\,s$^{-1}$, heating the  stellar ejecta to $\ \sim10^7$ K.
The main consequences of the ejecta-CSM interaction  is production of X-ray and radio emission  \citep{chevalier82b,cf03,cf16}.
An analytical model of the emission due to ejecta-CSM interaction in  SNe  was developed by \citet{chevalier82b}, in which ejecta moves freely for several years in the 
ejecta dominated phase (mass of the ejecta is larger than the swept-up CSM mass).
In normal core-collpase SNe,  it takes several thousands of years for the swept-up CSM mass to be significant enough (as compared to the ejecta mass)
to slow down the free moving ejecta. 
However, in SNe IIn, the extremely dense CSM can substantially decelerate the SN ejecta very quickly, and converts ejecta kinetic energy into radiation very efficiently. This also causes the  forward shock to cool radiatively and collapse into a cool
dense shell.

Since the shock velocities  are
typically 100--1000 times the speed of the progenitor winds, the shock
wave samples the wind lost many hundreds--thousands of years ago and thus  probes the
past history of the star.

When the interaction region between the forward and reverse shocks is treated in a thin-shell approximation, the  shock front evolution is characterized by a self-similar solution \citep{chevalier81, chevalier82a, chevalier82b}. By balancing  
the ram pressure from the CSM and the  ejecta, one can derive the  the shock radii $R_s$ expansion, which is  a power law in time as $R_s \propto t^m$ ($m=(n-3)/(n-2)$).  These solutions are valid for $n>5$ to meet the finite energy constraints  \citep{mm99}.

\citet{cf16} have shown how to obtain various shock parameters. Please refer to it for details. Here I write down the main equations needed to interpret the observational
data. Throughout the chapter, the  subscript `CS' refers to circumstellar shock and 'rev' to reverse shock.

The swept up masses behind the two shocks can be written as

\begin{eqnarray}
M_{\rm CS}=\frac {\dot M R_{\rm s}}{v_{\rm wind}}\\ \nonumber
M_{\rm rev}= 4 \pi \int_{R_s}^\infty \rho(r) r^2 dr
\label{eq:M}
\end{eqnarray}

 Since  $\rho_{\rm rev}/\rho_{\rm CS} = \rho_{\rm ej}/\rho_{\rm wind}$,  the 
masses and the densities of the shocked
shells are related as
\begin{eqnarray}
M_{\rm rev}=\frac{(n-4)}{2}M_{\rm CS}\\ \nonumber
\rho_{\rm rev}=\frac{(n-4)(n-3)}{2} \rho_{\rm CS}
\label{density}
\end{eqnarray}

The maximum ejecta velocity  $V_{\rm ej}$,
circumstellar shock velocity at the contact discontinuity  $V_{\rm s}$,  and reverse shock velocity $V_{\rm rev}$
 can be written as

\begin{eqnarray}
V_{\rm ej}=R_s/t \propto t^{-1/(n-2)}\\ \nonumber
V_{\rm s} =dR_s/dt=(n-3)/(n-2)V_{\rm ej}\\ \nonumber
V_{\rm rev}=V_{\rm ej}-V_{\rm s}=V_{\rm ej}/(n-2)
\label{eq:V}
\end{eqnarray}

 The temperature of the forward CSM shock $T_{\rm CS}$  (assuming 
cosmic abundances),  and that of the reverse shock $T_{\rm rev}$ are

\begin{eqnarray}
kT_{\rm CS}=117 \left(\frac{n-3}{n-2}\right)^2 \left(\frac{V_{\rm ej}}{10^4 \,\rm km \, s^{-1}}\right)^2 \rm keV\\ \nonumber
 kT_{\rm rev}=  \frac{T_{\rm CSM}}{(n-3)^2}=1.2 \left(\frac{10}{n-2}\right)^2 \left(\frac{V_{\rm ej}}{10^4 \,\rm km \, s^{-1}}\right)^2 \rm keV
 \label{eq:T}
\end{eqnarray}
The above equations  assume  energy  equipartition between ions and electrons,  for which the characteristic time is
\begin{equation}
t_{\rm eq}=2.5\times10^7 \left(\frac{T_e}{10^9\,\rm K}\right)^{1.5} \left(\frac{n_e}{10^7\, \rm cm^{-3}}\right)^{-1} \, \rm s
\label{teq}
\end{equation} 
The ratio of the electron temperatures of the shocked CSM and shocked ejecta is $(n -3)^2$.
Due to the  low temperature and higher density, the  condition of equipartition between ions and electrons 
may be valid in the reverse shock, but it is highly unlikely for the forward shock to obtain electron ion
equilibrium. However, in SNe IIn, it is possible to meet this condition for both reverse as well as forward shocks owing to their extreme densities.

\subsection{Radiation in ejecta CSM interaction}

During ejecta CSM interaction in SNe, radiation can come in multiple frequency bands from the shocked as well as unshocked regions via multiple channels. 

In the shocked CSM, the photospheric photons may undergo inverse-Compton (IC) scattering by the energetic electrons and may emit in UV and X-rays.
For an electron scattering optical depth $\tau_{\rm e}$ with $\tau_{\rm e} <1$ behind the CSM, a fraction $\tau_{\rm e}^N $ electrons will scatter $N$ times and boost their energies to
UV and X-ray with a power-law  spectrum of  spectral index between $-1$ to $-3$ \citep{cf03}.

In addition, forward and reverse shocks are very hot (Eq. \ref{eq:T}) and can emit soft and hard X-rays by free-free  bremsstrahlung radiation. 
The total X-ray luminosity in free-free emission
depends upon the medium density $\rho_i$  (where subscript $i$ refers to `CS' and `rev' for circumstellar and reverse shocks,
respectively), and the emitting volume, which is proportional to  $ R_s^3$.
 If $\Lambda$ is the cooling function such that $\Lambda \sim T^{0.5}$ \citep{cf94}, then the
X-ray luminosity 
can be written as 
$L_i\propto \rho_i^2 R_s^3 \Lambda $  \citep{flc96}. 
\citet{cf03} derives the expression for the free-free X-ray luminosity from the forward and reverse shocks  to be

\begin{equation}
L_{\rm i} \approx 3.0 \times 10^{39} \overline g_{\rm ff} C_{\rm n} \left( \frac{\dot M}{10^{-5} M_\odot\, \rm yr^{-1}} \right)^2  \left( \frac{v_{\rm wind}}{10\, \rm km\,s^{-1}} \right)^{-2}  \left( \frac{t}{10\, \rm day}\right)^{-1}
 \, \rm erg s^{-1}
 \label{eq:xray}
 \end{equation}
 Here $C_n$ depends on the ejecta density index, and is 1  for the forward shock  and  $(n - 3)(n -  4)^2/4(n - 2)$ for 
the reverse shock.     The $\overline g_{\rm ff}$ is the free-free Gaunt factor.
 A caveat to note  here is that this equation is valid only under electron-ion
equipartition which is questionable for the forward shock.

Eq. \ref{eq:xray} suggests that the free-free X-ray luminosity evolves as $t^{-1}$. However this is valid only for the constant mass loss  medium ($s=2$). For the general case, 
 \citet{flc96} derive the X-ray luminosity evolution to be  $$L_i \propto  t^{-(12-7s +2ns -3n)/(n-s )}$$
An important point is  that  this formula assumes the X-ray emission covers the full X-ray band, whereas the  X-ray satellites observe  only in a very narrow band, e.g. {\it  Chandra} and {\it XMM}-Newton cover 0.2--10~keV range.  This means unless the luminosity ratio in the narrow observing band is same as that in the total X-ray band, one will not see
the above evolution of the X-ray luminosity. 
As shown in \citet{flc96}, the luminosity in a given energy band with  $E < kT_i$ can be written as 
$L_i \propto t^{-(6-5s+2ns-3n)/(n-s)}$, which has a much flatter time dependence. For steady wind ($s = 2$), 
 $L_i \propto t^{-(n-4)/(n-2)}$.

The above treatment is valid for adiabatic shocks with $T >  2 \times 10^7$~K.  For high mass loss rates and slow moving shocks, the radiative cooling may be important
as the cooling function will depend on temperature as $\Lambda \propto  T^{-0.6}$ \citep{cf94}.
If the temperature of the shocks goes below $2 \times 10^7$~K, then the efficient radiative cooling will lower the ejecta temperature to
$10^4$\,K, where photoelectric heating from the shocks balances the cooling. Since the ram pressure is maintained, a cool dense shell (CDS) will form.
The column density of the CDS can be estimated from $N_{\rm cool} = M_{\rm rev}/4 \pi R_{\rm s}^2 m_p$ \citep[$m_p$ is mass of a proton; ][]{cf03}, which can be
simplified to
 
 \begin{equation}
 N_{\rm cool}({\rm RS}) \sim 10^{21} (n-4) \left( \frac{\dot M}{10^{-5} M_\odot\, \rm yr^{-1}} \right)  \left( \frac{v_{\rm wind}}{10\, \rm km\,s^{-1}} \right)^{-1}  \ \left( \frac{V_{\rm ej}}{10^4 \rm km\,s^{-1}}\right)^{-1}  \left( \frac{t}{100\, \rm days}\right)^{-1}\, \rm cm^{-2}\\
 \label{eq:column}
 \end{equation}
The effect of the CDS is that  it  will absorb  the soft X-rays  emission from the reverse  shock and re-radiate   into UV and optical emission and will contribute significantly
to bolometric luminosity \citep{fransson84}.
 In the CDS, the atoms recombine and produce mainly  H$\alpha$ emission lines of  intermediate line widths $\sim 1000$ km\,s$^{-1}$.

Even though the intrinsic X-ray luminosity is high from the reverse shock owing to its high density, due to absorption of the emission \citep{nymark+06}, 
the X-ray emission may be dominated 
by the forward shock, especially in SNe IIn \citep{chandra+12b,chandra+15}.  An additional factor favoring the dominance  of the  forward shock
is that once the reverse shock becomes radiative, its luminosity rises proportional to density while the luminosity from the adiabatic forward shock continues to grow as density squared.
If the mass loss rate is very high, even the forward shock may become radiative \citep[e.g. SN 2010jl,][]{chandra+15}.

While  the radiation due to ejecta-CSM  interaction in SNe is affected by complicated hydrodynamics, in a simple approach the radiation luminosity of the shocks  is the total kinetic luminosity times the radiation efficiency $\eta = t / (t + t_c )$, where $t_c$ is the cooling time of the shocked shell  at the age $t$.  In normal core-collapse SNe, the total radiated energy is only a few  per cent of its kinetic energy, but 
could reach up to   50\%  in SNe IIn  due to efficient cooling \citep{smith16a}.
In case of  radiative forward and reverse shocks, the kinetic  luminosity can be written as \citep{cf16}

\begin{equation}
L_i= \pi R_{\rm s}^2(1/2 \rho_i V_i^3)
= c_i  \frac{(n-3)}{(n-2)}  \times \frac{1}{2} \frac{\dot M V_{\rm ej}^3}{v_{\rm wind}}
\end{equation}
Here $c_i=(n-4)/2(n-2)^2$ for the reverse shock and $c_i=(n-3)^2/(n-2)^2$ for  the forward shock. Thus radiative luminosity goes as $L_i \propto t^{-3/(n-2)}$ for a steady
wind. However, for the  more general case, it will evolve as  $L_i \propto  t^{-(15-6s+ns-2n)/(n-s)}$.

While the X-ray emission  discussed above  is  from the shocked shells, the unshocked ejecta and CSM can also contribute to  significant radiation. 
During the shock breakout, the unshocked CSM can be ionized and emit in optical and UV bands by subsequent recombination, mainly in H$\alpha$,
Lyman-$\alpha$
and C III, C IV, N V, and Si IV, whose line widths gives information about the wind velocity. The unshocked CSM can also be excited by the X-rays from the forward
and reverse shocked shells  and can produce coronal emission lines, e.g., Fe VII and Fe X lines . The outermost layers of the unshocked ejecta, ionized and heated by radiation can contribute towards broad emission and absorption lines, such as C III, C IV, N V, and Si IV.
  In dense CSM,  electron scattering can be the  dominant scattering and broaden the width of the above discussed emission lines.

The above treatment assumes  spherical symmetry of the system. However, if the CSM  is asymmetric and/or clumpy, there are 
strong observational consequences for the CSM interaction with the SN ejecta \citep{cd94}.
In a clumpy CSM, the 
predominance of soft radiation is  likely to come out from the clumps,  since due to high density the velocity of the shock in the 
clump will be  much smaller than the  $V_c=V_s(\rho_s/\rho_c)^{1/2}$.  The higher density and lower velocity of the clump can lead to radiative cooling of the
clump shock. Indeed, if the bulk of the CSM were in clumps,
radiation from clumps become the dominant source of the radiation in  the soft XUV band.

 The ejecta diagnostics based on the CSM interaction cannot constrain the ejecta mass and energy uniquely because the same density versus velocity distribution in the ejecta outer layers can be produced by a different combination of mass and energy. Yet the observed interaction luminosity and the final velocity $V_f$ of the decelerated shell
 constrain the energy of the outer ejecta from the condition
 $V > V_f$. 
 For the  power-law index $n$ in $\rho_{\rm ej} \propto r^{-n}$,   the obvious relations, such as the ejecta density turnover velocity
 $V_0 \propto (E/M)^{1/2}$, $\rho (V) 
 \propto \rho_0 (V_0/V)^n$, and $\rho_0 \propto M/V^3$ 
 result in the energy-mass scaling $E \propto M^{(n-5)/(n-3)}$.
This scaling, when combined with the requirement   that the velocity $V_f$ should be larger than  $V_0$, provides us with the lowest plausible values of 
 $E$ and $M$. 

An important consequence of ejecta-CSM interaction is the  radio emission. In the fast moving forward shock, the 
particles can be accelerated to relativistic energies,  and  the presence of  magnetic field give rise to non-thermal 
synchrotron radio emission, which we discuss below.

\subsubsection{Radio emission}

In the shock front, the charged particles are accelerated most likely via diffusive Fermi shock acceleration \citep{bo78,bell78}.
The magnetic field is most likely the  seed CSM magnetic field, compressed and enhanced in the shocked region.  Rayleigh-Taylor instabilities at the contact 
discontinuity can further enhance the magnetic field \citep{chevalierblondin95}.
If the energy density of the magnetic field as well as the relativistic particles are considered to be proportional to the thermal energy density, then one can easily obtain the
radio emission formulae.

 Radio emission arises  primarily from the higher temperature forward shock, where it is easier to accelerate the particles to relativistic energies. The electron energy distribution is  assumed to
be a power law  $N(E) = N_o E^{-p}$ , where $E$ is the electron energy, $N_o$ is the normalization of the distribution and $p$ is the electron energy index,
which is related to spectral index $\alpha$ (in $F_\nu \propto \nu^{-\alpha}$, where $F_\nu$ is the radio flux density at frequency $\nu$) as
$\alpha=(p-1)/2$. 

The radio emission is affected by either the external free-free absorption (FFA) process by the surrounding ionized wind, or by the internal synchrotron self absorption (SSA)
by the same electrons responsible for the emission.  The dominant absorption mechanism depends upon the mass loss rate, magnetic field
in the shocked shells, shock velocity and density of the ejecta. One can distinguish between the two processes from the optically thick part of the light curve or
spectrum.
 
 If SSA is the  dominant absorption mechanism, then for  SSA optical depth $\tau_\nu^{\rm SSA}$,   one can use the \citet{chevalier98} formulation to derive radio flux density evolution.

\begin{eqnarray}
F_\nu^{\rm SSA} =  \frac{\pi R_s^2}{D^2} \frac{c_5}{c_6} B^{-1/2} \left(\frac{\nu}{2c_1}\right)^{5/2}  [1-\exp(-\tau_\nu^{\rm SSA})]\\ \nonumber
\tau_\nu^{\rm SSA}= \int\limits_{0}^{s} \kappa(\nu)ds \approx \kappa(\nu) s =   \left(\frac{\nu}{2c_1}\right)^{-\frac{p+4}{2}}  \left( \frac{4fR_sc_6N_o B^{\frac{p+2}{2}}}{3}\right)
\label{SSA-flux}
\end{eqnarray}
where  $c_1$, $c_5$ and $c_6$ are the constants defined in \citet{pacholczyk70},  $D$, and $f$, and   $B$ are the  distance to the SN from observer, filling factor and magnetic field strength, respectively.

One can determine $N_o$ from $\int_{E_l}^\infty N(E) E dE$,  assuming electron rest mass energy to be the lower energy limit, i.e. $E_l=0.51$\,MeV, then 
$$ N_o=\frac{a B^2 (p-2) E_l^{p-2}}{8 \pi}$$
where $a$ is the equipartition factor.
The above equation can be written in terms of optically thick $\tau>1$ and optically thin $\tau \le 1$  limits as

\begin{eqnarray}
F_\nu^{\rm SSA}|_{\tau \le1}=\frac{4\pi f R^3}{3D^2}c_5 N_o B^{\frac{p+1}{2}}  \left(\frac{\nu}{2c_1}\right)^{-\frac{p-1}{2}} \\ \nonumber
F_\nu^{\rm SSA}|_{\tau >1}= \frac{\pi R^2}{D^2} \frac{c_5}{c_6} B^{-1/2} \left(\frac{\nu}{2c_1}\right)^{5/2} 
\label{thinthick}
\end{eqnarray}

From these equations, one can estimate the $R_p$ and $B_p$ at the peak of the spectrum, when $F_\nu^{\rm SSA}$ is $F_p$. 
They will depend on various parameters as
\begin{eqnarray}
R_p \propto a^{-\frac{1}{2p+13}} f^{-\frac{1}{2p+13}} (p-2)^{-\frac{1}{2p+13}} F_p^{\frac{p+6}{2p+13}} D^{\frac{2p+12}{2p+13}} \nu^{-1}\\ \nonumber
B_p \propto a^{-\frac{4}{2p+13}} f^{-\frac{4}{2p+13}} (p-2)^{-\frac{4}{2p+13}} F_p^{-\frac{2}{2p+13}} D^{-\frac{4}{2p+13}} \nu
\label{thinthick2}
\end{eqnarray}
\citet{chevalier98} has given these equations for $p=3$ to be:
\begin{eqnarray}
R_p =8.8\times10^{15} a^{-\frac{1}{19}} \left( \frac{f}{0.5} \right)^{-\frac{1}{19} }  \left( \frac{F_p}{\rm Jy} \right)^{\frac{9}{19}}  \left( \frac{D}{\rm Mpc} \right)^{\frac{18}{19}}  \left( \frac{\nu}{5\, \rm GHz} \right)^{-1}\rm cm\\  \nonumber
B_p = 0.58 a^{-\frac{4}{19}} \left( \frac{f}{0.5} \right)^{-\frac{4}{19}}   \left( \frac{F_p}{\rm Jy} \right)^{-\frac{2}{19} } \left( \frac{D}{\rm Mpc} \right)^{-\frac{4}{19} } \left( \frac{\nu}{5\, \rm GHz} \right)\, G\rm 
\label{thinthick3}
\end{eqnarray}

In Eq. \ref{thinthick2}, 
if $R$ can be measured in some independent way (e.g. VLBI), then both the magnetic field and the column density of the relativistic electrons can be determined.
Another possibility to independently estimate the magnetic field is from synchrotron cooling effects, if it is observationally seen in the supernova spectrum. 
One can derive the exact equipartition fraction \citep[e.g. SN 1993J,][]{chandra04}.

 By relating the post-shock magnetic energy density to the shock ram pressure, i.e. $B^2/8 \pi= \varsigma \rho_{\rm wind}  V_s^2 $, where $ \varsigma$ ($ \varsigma \le 1$)
is a numerical constant,  mass loss rate can be obtained as

\begin{equation}
\dot M = \frac{6\times10^{-7}}{m_H^2 \varsigma} \left( \frac{B}{1\, \rm G} \right)^2 \left( \frac{t}{100\, \rm d} \right)^2 \left( \frac{v_{\rm wind}}{10\, \rm km\,s^{-1}} \right) \rm M_\odot yr^{-1}
\end{equation}

While SSA seems to dominate in high velocity shocks, in the case of high mass loss rate, external free-free absorption may be quite dominant. 
The optical depth of the free-free absorption can be defined as 

$$\tau_\nu^{\rm FFA}=\int\limits_{R_{\rm S}}^{\infty} \kappa_\nu^{\rm FFA} n_e n_i ds$$
Here $\kappa_\nu^{\rm FFA}$  is the absorption coefficient, which depends upon $n_e$ and $n_i$, the electron and ion number densities, respectively. 
Here $n_e=\dot M/4\pi R_s^2 v_{\rm wind} \mu_e m_H$, and $n_i=\overline Z n_e$, where $\overline Z=\sum X_j Z_j^2/ \sum X_j Z_j$.
Thus
\begin{equation}
\tau_\nu^{\rm FFA}=\frac{\dot M \overline Z \kappa_\nu^{\rm FFA} }{3 (4\pi)^2 R^3 m_H^2 \mu_e^2 v_{\rm wind}^2}
\end{equation}
\citet{pf75, weiler+02} derived $\kappa_\nu^{\rm FFA}$ to be
\begin{equation}
\kappa_\nu^{\rm FFA}=4.74\times10^{-27} \left(\frac{\nu}{1\, \rm GHz}\right )^{-2.1} \left( \frac{T_e}{10^5 \, \rm K} \right)^{-1.35}
\end{equation}
FFA will cause the radio emission arising out of the (mainly) forward shock to be attenuated by $\exp(-\tau_\nu^{\rm FFA})$.

The above equation for a fully ionized wind, and assuming the wind
 to be singly ionized, i.e. $\overline Z=1$ and $\mu_e=1.3$, 
gives a mass loss rate 

\begin{equation}
\dot M =4.76\times10^{-5}  (\tau_{\nu}^{\rm FFA})^{0.5}\left( \frac{V_{\rm ej}}{10^4\, \rm km\,s^{-1}} \right)^{1.5} \left( \frac{t}{100\, \rm d} \right)^{1.5} 
\left( \frac{T_e}{10^5\, \rm K} \right)^{0.675}  \left( \frac{v_{\rm wind}}{10\, \rm km\,s^{-1}} \right) \rm M_\odot yr^{-1}
\end{equation}
Here it is important to mention that the above formula will overestimate the mass loss rate if the wind if clumpy. 
\citet{puls+08} has shown that in a clumpy wind, the mass loss rate can change (to a lower value) by as much as a factor of 3.

In addition to SSA and FFA, \citet{weiler+90} proposed a model in which thermal absorbing gas is mixed into the synchrotron emitting gas. This can be a likely case with
SNe IIn, in which a CDS is formed. A fraction of the cool gas ($10^4-10^5$\,K) from the CDS mixed in the forward shocked region can also give rise to internal FFA.
This will attenuate the radio emission by a factor $(1-\exp(-\tau_\nu^{\rm intFFA}))$, where  $\tau_\nu^{\rm intFFA}$ is the internal FFA optical depth.

To fit the observational data, one can put  the above formulae in simpler forms , and derive time and
frequency dependence of various parameters.
When SSA is the dominant absorption mechanism, the data can be fit with the following model, 

\begin{eqnarray}
F_\nu (t)=K_1 \nu^{5/2} t^{a} (1-\exp(-\tau_\nu^{\rm SSA}))\\ \nonumber
\tau_\nu^{\rm SSA}=K_2 \nu^{-(p+4)/2} t^{-(a+b)},
\end{eqnarray}
Here $a$ gives the time evolution of the radio flux density in 
  the optically thick phase ($F \propto t^a$) and $b$ in the optically thin phase ($F \propto t^{-b} $). Under the assumption that the energy density in the particles and the fields is proportional to the postshock energy density, these quantities are related with expansion parameter $m$ and electron energy index $p$ as $a = 2m + 0.5$, and 
$b = (p + 5 - 6m)/2$.

When FFA is the dominant absorption mechanism, the evolution of the radio flux density can be fit with

\begin{eqnarray}
F_\nu (t)=K_1 \nu^{-\alpha} t^{-\beta} \exp(-\tau_\nu^{\rm FFA}(t))\\ \nonumber
\tau_\nu^{\rm FFA}(t)=K_2 \nu^{-2.1} t^{-\delta},
\end{eqnarray}
where $\alpha$ is the frequency spectral index, which relates to the electron energy index  as $p=2\alpha+1$.
Here $K_1$ is the radio flux density normalization parameter and $K_2$ is the FFA optical depth normalization parameter. The parameter  $\delta$ 
is related to the expansion parameter $m$ in $ R_s \propto t^m$ as $\delta \approx 3m$.
The parameter $\beta$ is the time dependence, which under the  assumption that the energy density in the particles and the fields is proportional to the postshock energy density leads to $\beta = (p + 5 - 6m)/2$ \citep{chevalier82a,chevalier82b}. 

When the  mixing of cool gas in the synchrotron emitting region is responsible for much of the absorption, in such a case the flux takes the form

\begin{eqnarray}
F_\nu (t)=K_1 \nu^{\alpha} t^{\beta} \frac{(1-\exp(-\tau_\nu^{\rm intFFA}))}{\tau_\nu^{\rm intFFA}}\\ \nonumber
\tau_\nu^{\rm intFFA}=K_2 \nu^{-2.1} t^{\delta'},
\end{eqnarray}
where $\delta'$ is time evolution of internal-FFA optical depth.

The above discussion is based on the  spherical geometry of the SN-CSM system.  However, there is evidence that   the ejecta and the CSM may be far from spherical and very  complex in some cases. \citet{smith+09} found  an anisotropic and clumpy CSM medium in the  RSG VY CMa, whereas the structure of $\eta$ Carinae was
found to be bipolar \citep{smith10}.
The deviation from spherical geometry  can have strong observational consequences in studying the SN-CSM interaction.

From this section onwards, I will concentrate on the observational aspects of SNe IIn, mainly in X-ray and radio bands.

\section{X-ray observations of SNe IIn}
\label{sec:observations}

\begin{figure}
\begin{center}
 \includegraphics[angle=0,width=0.98\textwidth]{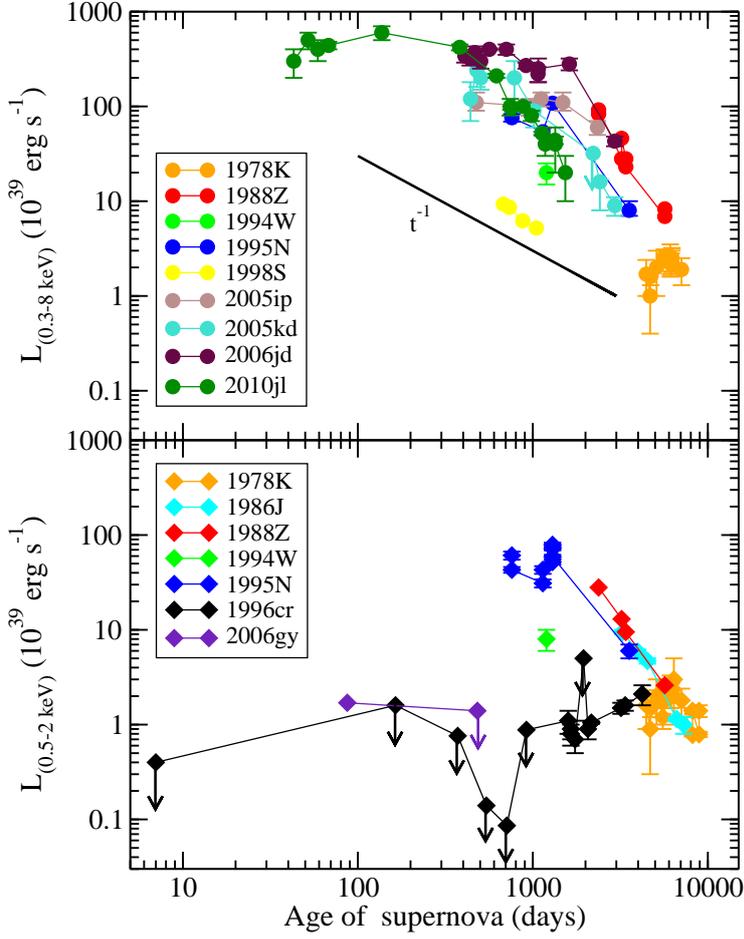}
 \caption{The 0.3--8 keV (upper panel) and 0.5--2 keV (lower panel) X-ray luminosities of SNe IIn \citep{rd17}.
In the upper panel the X-ray luminosities of SN 1998S and SN 2006jd are in 0.2-10\,keV and in the lower panel,  SN 1986J and SN 1995N X-ray luminosities 
are in the range 0.5--2.4\,keV.}
   \label{fig:xray}
\end{center}
\end{figure}

Since a high CSM density is a prerequisite for the radio and X-ray emission, 
 the ejecta CSM interaction is expected to emit copiously in these  bands in SNe IIn. 
However, the statistics are contradictory. While  more than $\sim 400$ SNe IIn are known, only 12 are known to emit in the X-ray bands (Fig. \ref{fig:xray}). 
These include SNe 1978K, 1986J, 1988Z, 1995N, 1994W, 1996cr, 1998S, 2005ip, 2005kd, 2006gy, 2006jd and 2010jl \citep{rd17}. In Fig. \ref{fig:xray},
we plot X-ray luminosities of SNe IIn in 0.3--8~keV and 0.5--2~keV bands.  In 0.3--8~keV band, SNe IIn have high luminosities spread within an
order of magnitude, with the exception of SNe 1978K and 1998S which are weaker X-ray emitters. Other than SN 2010jl, most of the X-ray detected SNe IIn have X-ray emission after around a year. This could have contribution from  observational biases due to the lack of early observations. However, recently,  in SN 2017gas,  a nearby ($d=43$\,Mpc)
SN IIn, early observations  covering 20--60 days after  the discovery resulted  in a non-detection \citep{cc17}.
In lower panel of Fig. \ref{fig:xray}, we plot X-ray luminosity of SNe IIn in soft X-ray band (0.5--2\,keV).  SN 2006gy is a weak X-ray emitter emitting only in the soft X-ray band. However, it was a SLSN and SLSNe are generally known to be weak X-ray emitters.
Here the most unique SN is SN 1996cr, which was detected after 1000 days and since then it's luminosity continued to rise.

\begin{figure}
\begin{center}
 \includegraphics[angle=-90,width=0.98\textwidth]{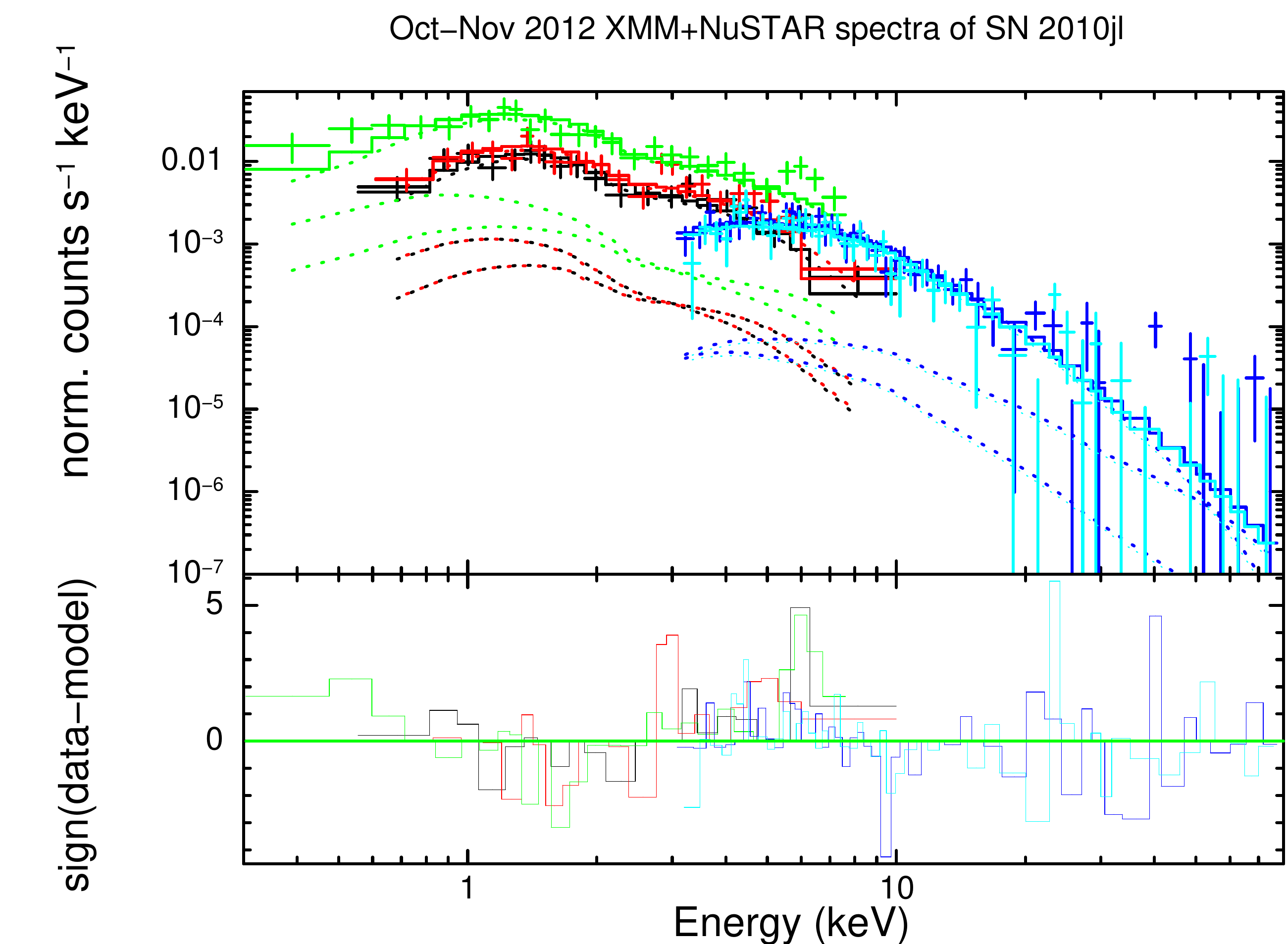} 
 \caption{Joint \textit{NuSTAR} and \textit{XMM-Newton} spectrum of the SN 2010jl covering energy range 0.3--80~keV. \citet[reproduced from,][]{chandra+15}.}
   \label{nustar}
\end{center}
\end{figure}

Due to higher density of the reverse shock, the thermal X-ray emission from core collapse SNe, in general, has a dominant contribution from the reverse shock.
In SNe IIn, the X-ray plasma temperatures have been found to be generally higher, consistent with the emission from the forward shock. In SN 2005ip, the 
X-rays were fit by temperature $T \ge 7$ keV, for at least upto 6 years after the explosion \citep{katsuda+14}. In SN 2005kd, the
early emission  (up to $\le2$ years) was dominated by hard X-rays  \citep{dwarkadas+16}.  \citet{chandra+12b} have shown that the {\it Chandra} and {\it XMM-Newton} observations in SN 2006jd are best fit with a thermal plasma with
electron temperature $\ge 20$\,keV.  This is because the 
radiative cooling shell (CDS)  formed between the reverse and forward shock may have absorbed  most of the X-rays emanating from the reverse shock.
While this implies X-ray emission is mainly coming from the forward shock,  these telescopes did not have the sensitivity to better constrain the temperature 
since both  work below 10\,keV range. In SN 2010jl, using the joint {\it NuSTAR} and {\it XMM-Newton} observations,  \citet{ofek+14b, chandra+15} derived the
 X-ray spectrum  in the energy range 0.3--80\,keV, and constrained the X-ray emitting shock  temperature to be 19\,keV, thus confirming that the dominant
X-ray emission in the SN is indeed coming from the hotter forward  shock (Fig. \ref{nustar}).
 This is the first time  the (forward) shock temperature was observationally accurately measured in a SNe IIn.

\begin{figure}
\begin{center}
 \includegraphics[angle=0,width=0.98\textwidth]{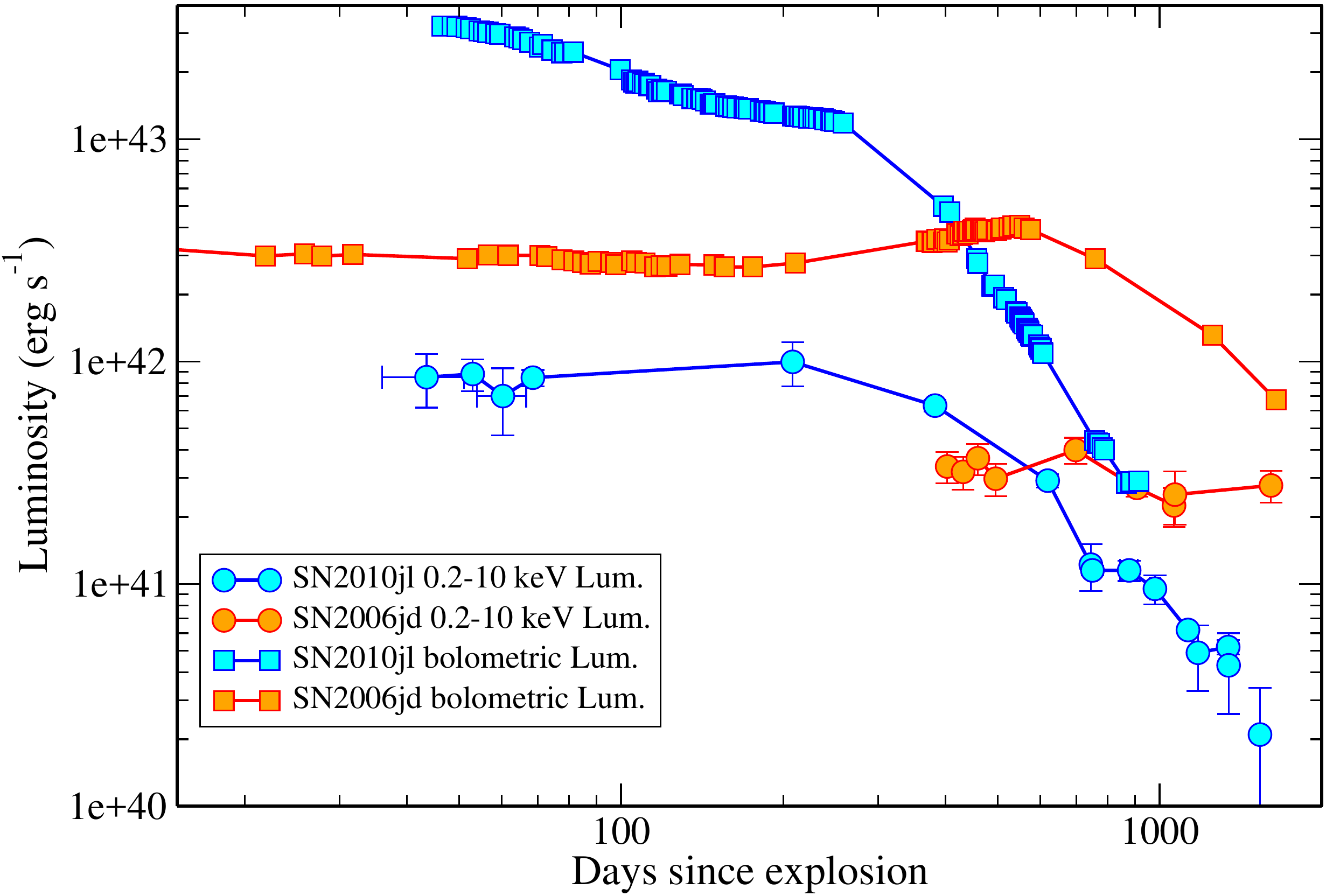}
  \caption{The X-ray (filled circles) and bolometric luminosity (filled squares) evolution of SN 2006jd and SN 2010jl. The blue curves are for SN 2010jl and orange curves indicate SN 2006jd. None of them follow $\sim 1/t$ dependence. \citet[reproduced from,][]{chandra+15}.}
    \label{fig3}
\end{center}
\end{figure}

As discussed in  section, \S \ref{CSM}, the thermal X-ray luminosity is expected to decline as $\sim 1/t$ for a steady wind. However, observations of 
several SNe IIn have shown to not follow this evolution 
 (See Fig.  \ref{fig:xray} as well as \citet{dwarkadas12, dwarkadas+16}). 
  In the energy range $0.2-10$~keV, where the current telescopes {\it Chandra},  {\it XMM-Newton} and {\it Swift-XRT} are most  sensitive,  SN 2005kd  followed a
 X-ray  luminosity evolution of $L_{\rm Xray} \propto t^{-1.6}$  \citep{dwarkadas+16}. Contrary to it,  SN 2006jd followed a  flatter decline of $t^{-0.2}$ for upto 4.5 years \citep{chandra+12b}, followed by a much steeper decline 8 years later \citep{katsuda+16}.
The X-ray luminosity for SN 2010jl was found to be roughly constant for the first $\sim 200$ days with a power-law index of $t^{0.13 \pm 0.08}$, followed by a
much faster decay with a power-law index of $\sim -2.12 \pm 0.13$ after day 400. Unfortunately there are no data between 200 and 400 days to see the luminosity evolution
in this range. 
Figure \ref{fig3} shows the plot of X-ray and bolometric luminosity light curves of SN 2010jl  and SN 2006jd. 
The deviation from $t^{-1}$ is quite evident here.  The luminosity decline is much flatter in SN 2006jd than that of SN 2010jl. Around the same epoch, the SN 2006jd X-ray light curve declines as $t^{-0.24}$, while SN 2010jl declines as $t^{-2.12}$. The relative flatness of the bolometric luminosity of SN 2006jd for a longer duration indicates that the CSM interaction powered the light curve for a much longer time than SN 2010jl. This indicates that the duration of mass ejection in SN 2006jd may have been longer in this case than for SN 2010jl, though in both cases it occurred shortly before the explosion. This suggests a different nature of progenitors for the two SNe.

 While this deviation from $L_{\rm Xray} \propto 1/t$ could be attributed to effects like  like inverse-Compton scattering etc. at early times, the late time deviation seen in many SNe IIn cannot be explained by it.
The complex nature of the progenitor star of SNe IIn could also be a cause of this. 
If the winds are not steady (deviation from $\rho_{\rm wind} \propto 1/R^2$ ), and have clumpy, asymmetric structure, the luminosity evolution will be far from $L_{\rm Xray} \propto 1/t$. However, the narrow energy ranges of 
present X-ray instruments pose a problem because  one measures spectral X-ray luminosity, whereas the   $\sim 1/t$ dependence  is valid for total X-ray luminosity \citep{flc96}.
As discussed above, the observations have indicated that the temperatures of the X-ray emitting regions in many SNe IIn are higher than can be measured in the 0.2--10 keV bandpass of the X-ray detectors,
thus we may be missing the peak of the X-ray emission.
A measurable effect of this instrument bias is that, as shock sweeps up more and more material with time, the shock temperature decreases, shifting the X-ray emission to progressively lower temperatures at
later epochs. Thus one should see flattening or even increase in the soft X-ray emission flux with time. 
This was seen in SN 1978K, where  2--10 keV flux showed 
decline, but 0.5--2\,keV soft X-ray emission remained constant for a long time without any signs of decline  \citep{schlegel+04}. However, there is are exceptions like  SN 1986J
and SN 1988Z
where the soft X-ray luminosity was found to evolve  as $L_{\rm Xray} \propto t^{-3}$ \citep{temple+05} and $L_{\rm Xray} \propto t^{-2.6}$ \citep{schlegel+06}, respectively. This may suggest that more physical reasons (like complex progenitor, unsteady mass loss rates etc.) are responsible for this behaviour, other than
the narrow bandpass of X-ray instruments. 
In this respect, SN 1996cr was a unique SN which 
demonstrated a factor of $>30$ increase in X-ray flux between 1997 and 2000 \citep{bauer+08}.
The observations implied that  the progenitor of SN 1996cr exploded in a cavity and freely expanded for 1--2 before striking the dense CSM.
 The above observational study shows that it is crucial to have wide band X-ray coverage to disentangle the instrumental issues from the physical reasons in order
 to truly understand the nature of the 
X-ray emission.

\subsection{Evolution of column density}

In some SNe IIn with well sampled X-ray observations, the X-ray column densities have seemed to evolve with time, indicating the cause of absorption as CSM 
as opposed  to ISM.
\citet{katsuda+14} found column density of $N_H \sim 5 \times 10^{22}$ cm$^{-2}$ for first few years in SN 2005ip, gradually decreasing to $N_H \sim 4 \times 10^{20}$ 
cm$^{-2}$ at later epochs, consistent with the Galactic absorption. The fact that the spectra was fit well with a thermal emission model with $kT > 7$ keV implied
a  forward shock origin of the 
X-rays  absorbed by the evolving CSM. 

\begin{figure}
\begin{center}
 \includegraphics[width=0.98\textwidth]{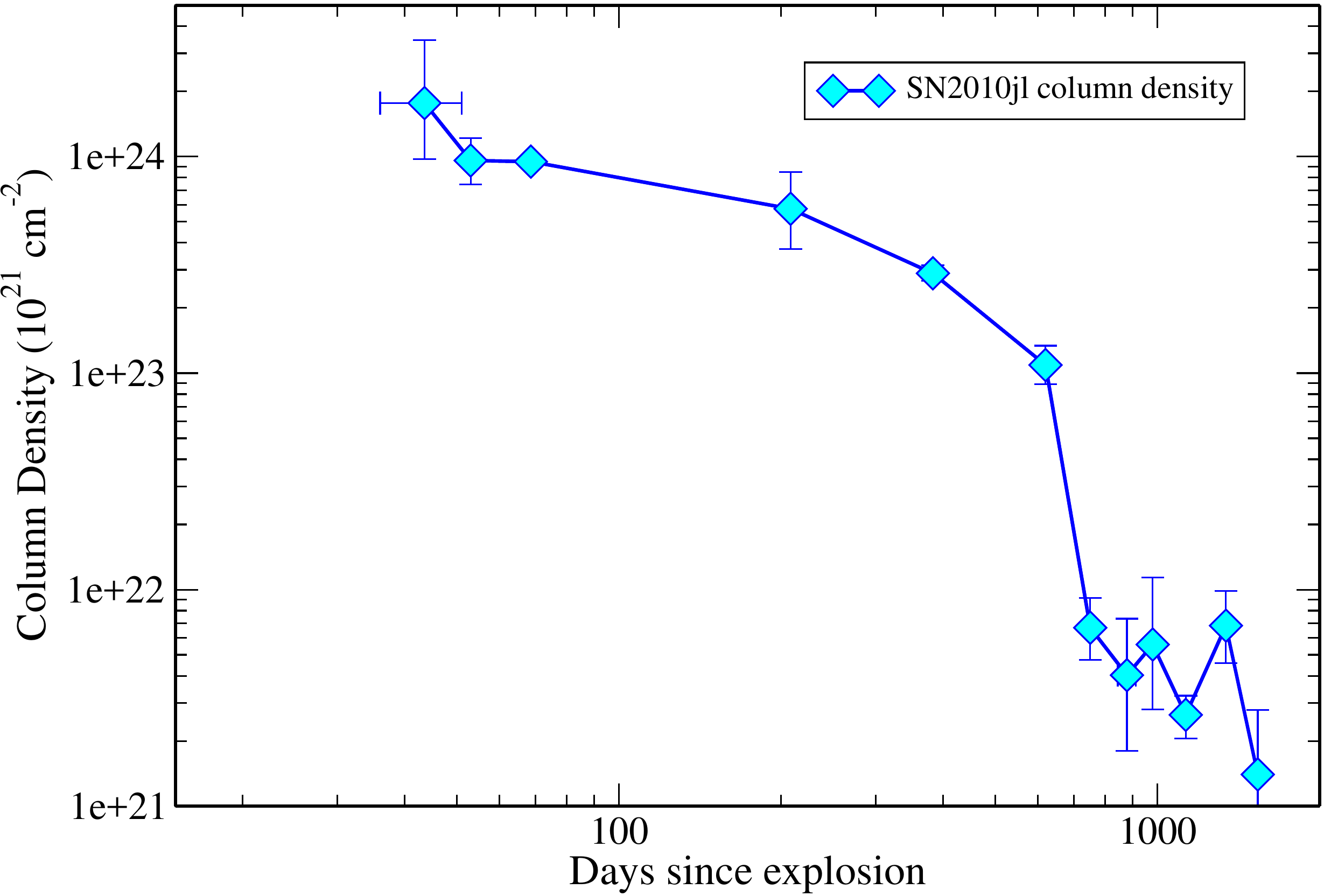} 
 \caption{The column density evolution in SN 2010jl. The figure is reproduced from \cite{chandra+15}.}
   \label{fig5}
\end{center}
\end{figure}

SN 2010jl is the best studied  Type IIn SN in X-ray bands, where X-ray observations covered a period up to day 1500 with  \textit{Chandra}, \textit{XMM-Newton}, \textit{NuSTAR}, and 
\textit{Swift}-XRT.
This SN provided the best studied evolution of column density  from 40 to 1500 days \citep[Fig. \ref{fig5} and][]{chandra+12a, chandra+15}. During this time period,  three orders of magnitude change in column density was witnessed.
At the first epoch the  column density associated with the SN,$N_H=10^{24}$ cm$^{-2}$,  which was 3000 times higher than the Galactic column density, which declined slowly by an order of magnitude
upto day $\sim650$. Followed by a  sharp decline, it settled to 10 times the Galactic value.  The higher column density  observed in the X-ray observations for SN 2010jl was not found to be associated with the high host galaxy extinction, indicating  that the higher column density is due to the mass loss near the forward shock  and thus is arising from the CSM. 
This was the first time that external circumstellar X-ray absorption had been clearly observed in a SN and enabled one to trace the precise mass loss evolution history of the  star.

As discussed in previous section, due to high density of the CSM, much of the X-ray is reprocessed to visible bands.  Thus one would expect less efficient reprocessing as
material becomes more transparent with time.
 In the  X-ray and bolometric luminosities plot of SN 2006jd and SN 2010jl (Fig. \ref{fig3}),  the bolometric luminoisities are at least an order of magnitude higher for first several hundred days. However,   the difference between the two luminosities is decreasing with time. This indicates that with decreasing density, reprocessing of X-rays is becoming less efficient. Long term follow up of the X-ray and bolometric 
luminosities will be very informative here.

\section{Radio emission of SNe IIn}
\label{sec:radio}

\begin{figure}
\begin{center}
 \includegraphics[angle=0,width=0.98\textwidth]{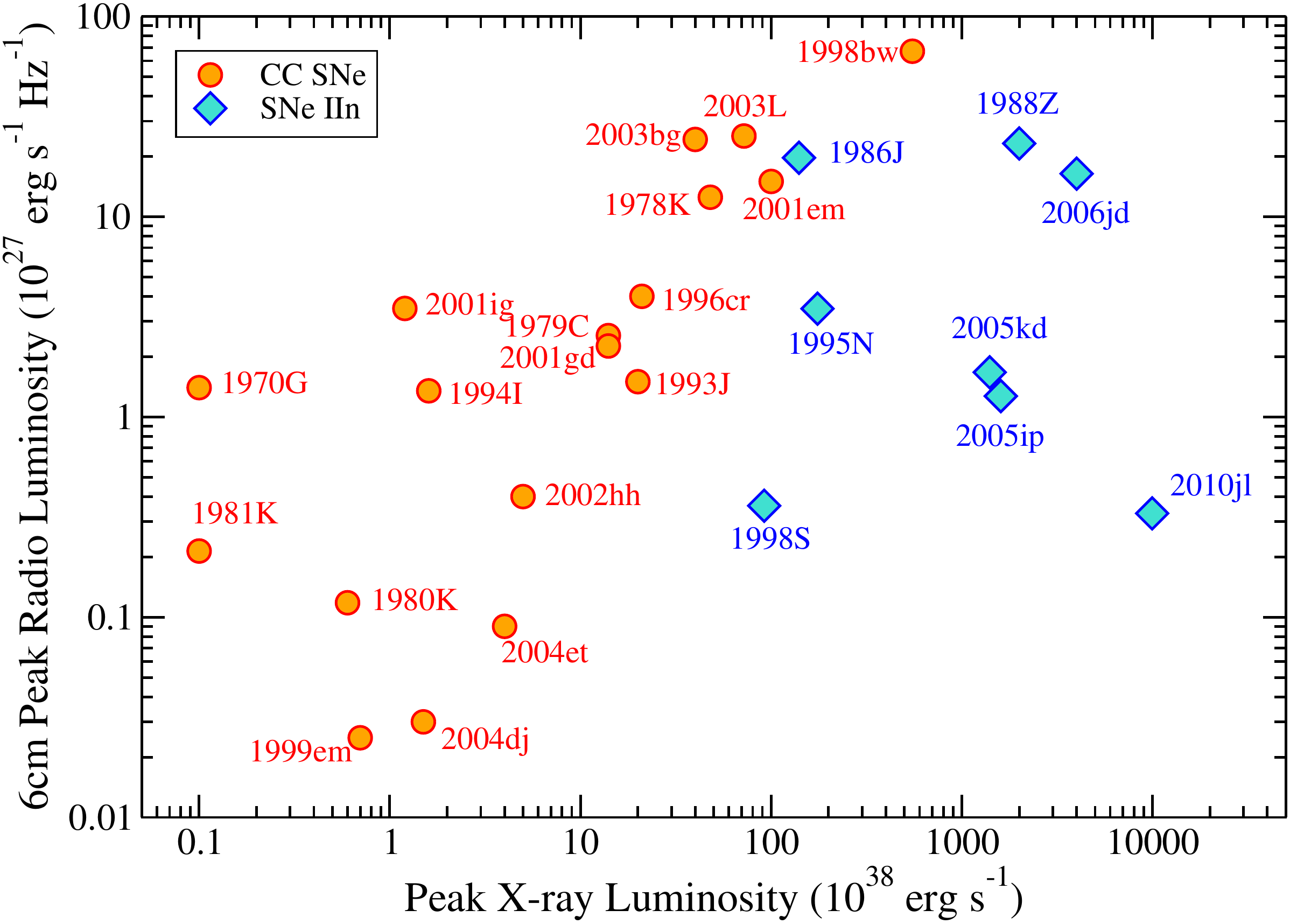}
 \caption{Radio and X-ray peak luminosities of SNe IIn compared with some well observed core collapse SNe. While in radio bands SNe IIn do not stand out, they
 occupy higher values in the X-ray luminosity space. The X-ray data are taken from \citet{immler07} and radio data  mainly from \citet{weiler+02}. }
   \label{fig2a}
\end{center}
\end{figure}

Radio emission in SNe IIn is very intriguing. 
 Out of $\sim 400$, less than half SNe IIn ($\sim 154$) have been looked at in the radio bands (mainly with the Very
Large Array (VLA)) and $\sim 10$\% have been detected. 
However, there is a very interesting trend in radio and X-ray detected SNe IIn. 
While X-ray luminosities are  towards the high end in SNe IIn, the radio luminosities are diverse. In  Fig. \ref{fig2a}, we plot peak X-ray and
radio luminosities of radio and X-ray detected SNe IIn and compare them with some well sampled core collapse SNe.  While the X-ray luminosities of
SNe IIn are systematically higher than their counterparts, radio luminosities do not stand out and occupy four  orders of magnitude space.

\begin{figure}
\begin{center}
   \includegraphics[angle=0,width=0.98\textwidth]{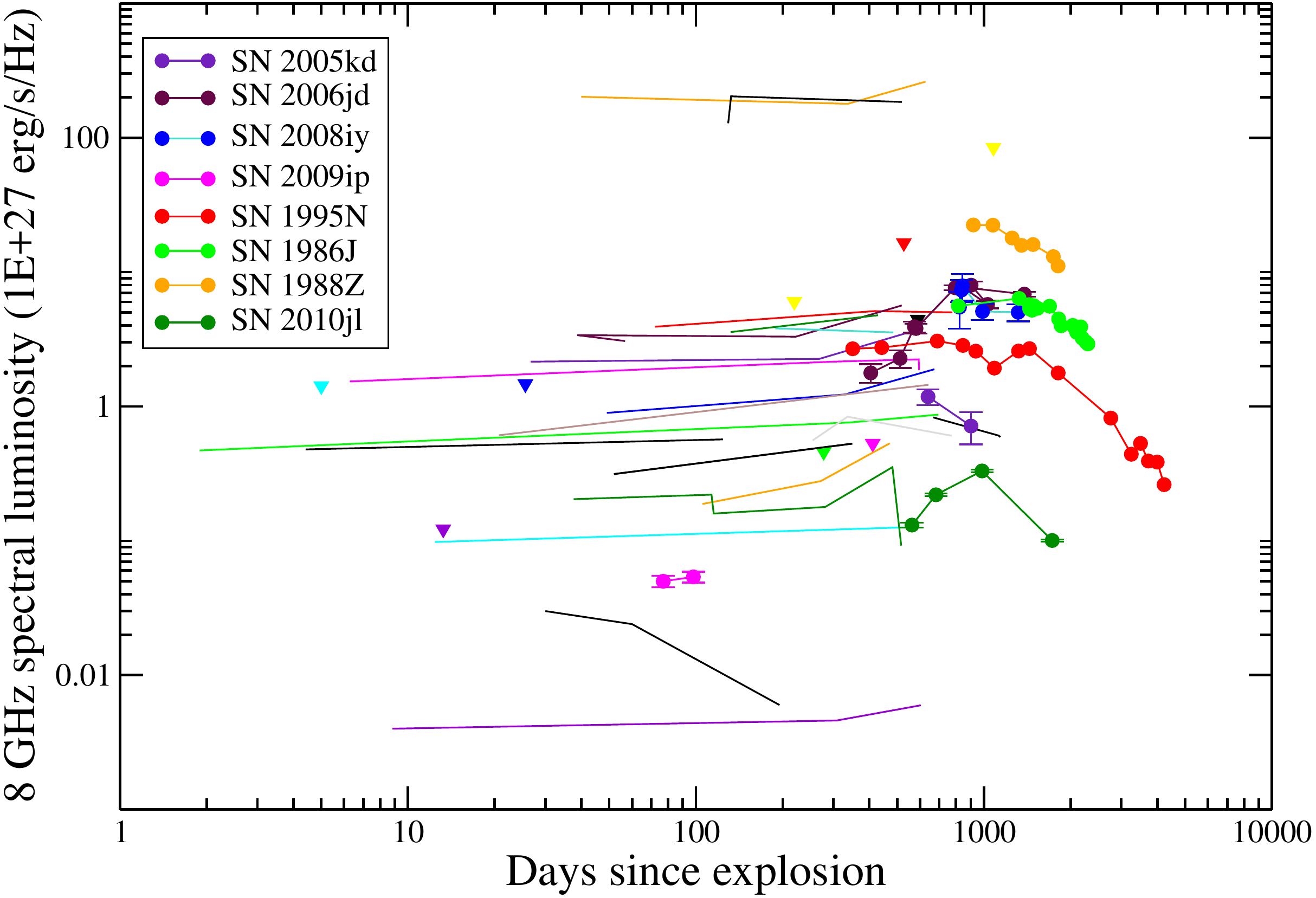}
 \caption{8\,GHz radio spectral luminosity of some of the well observed SNe IIn (Chandra et al. to be submitted). Here circles are detections. The lines indicates upper limits at
 observations at many epochs. Single triangles indicate upper limits at individual epoch. }
   \label{fig2}
\end{center}
\end{figure} 

This probably could be explained as a side effect of high density.
While high density  of the
CSM means efficient production of synchrotron radio emission, 
it also translates into higher
absorption. Hence, by the time emission reaches optical depth of unity, the strength of synchrotron emission is already weakened.
This can explain that radio  light curves are extremely diverse;  the diversity translate to diversity in the CSM density.

The support for  above argument also comes from the fact that 
most of the SNe IIn are late radio emitters (Fig. \ref{fig2}), indicating higher absorption early on (Chandra et al., to be submitted).
   Some of the radio emitting SNe IIn were either
discovered late or classified late. The prototypical SNe IIn, e.g. SN 1986J, SN 1988Z and SN 1978K were observed years after their explosions. 
Most of the   SNe IIn 
discovered soon after the explosion,  did not have early radio
observations within a month, except for SN 2009ip \citep{margutti+14}. SN 2009ip
was detected in radio bands early on but  faded below detection within a few tens of days. 
However, SN 2009ip was a  peculiar SN with  a poorly understood explosion mechanism. In contrast, the observations of SN 2010jl started by day 45 but the first detection happened
only after 500 days \citep{chandra+15}, clearly showcasing the efficient absorption of radio emission for a long period.

In addition to  high absorption, observational biases  could also be partially responsible. 
The late radio turn on may also explain the observed 
overall low detection rates of SNe IIn. If a SN is not bright in radio bands in a first few epochs, the observational campaign for that SN is usually over. So even  if a SN IIn was a potential radio emitter at later epochs, it would be missed in radio bands. To understand this observational bias, 
 \cite{vandyk+96} observed 10 SNe IIn at an age of a few hundred days and
still did not detect any; they set  upper limits of $150-250$ $\mu$Jy. However, the current telescopes are able to reach at least an order of magnitude better sensitivity. 
In addition, the current  upgraded VLA is extremely sensitive at higher frequencies where the radio absorption effects vanish much sooner ($\tau_\nu \propto \nu^{-2}$). 
Strategic study of SNe IIn at higher frequencies at early times to low frequencies at late epochs is 
likely to result in the most complete sample of radio supernovae \citep{pt+15,wang+15}.

\subsection{Absorption of radio emission}

\begin{figure}
\begin{center}
 \includegraphics[angle=0,width=0.98\textwidth]{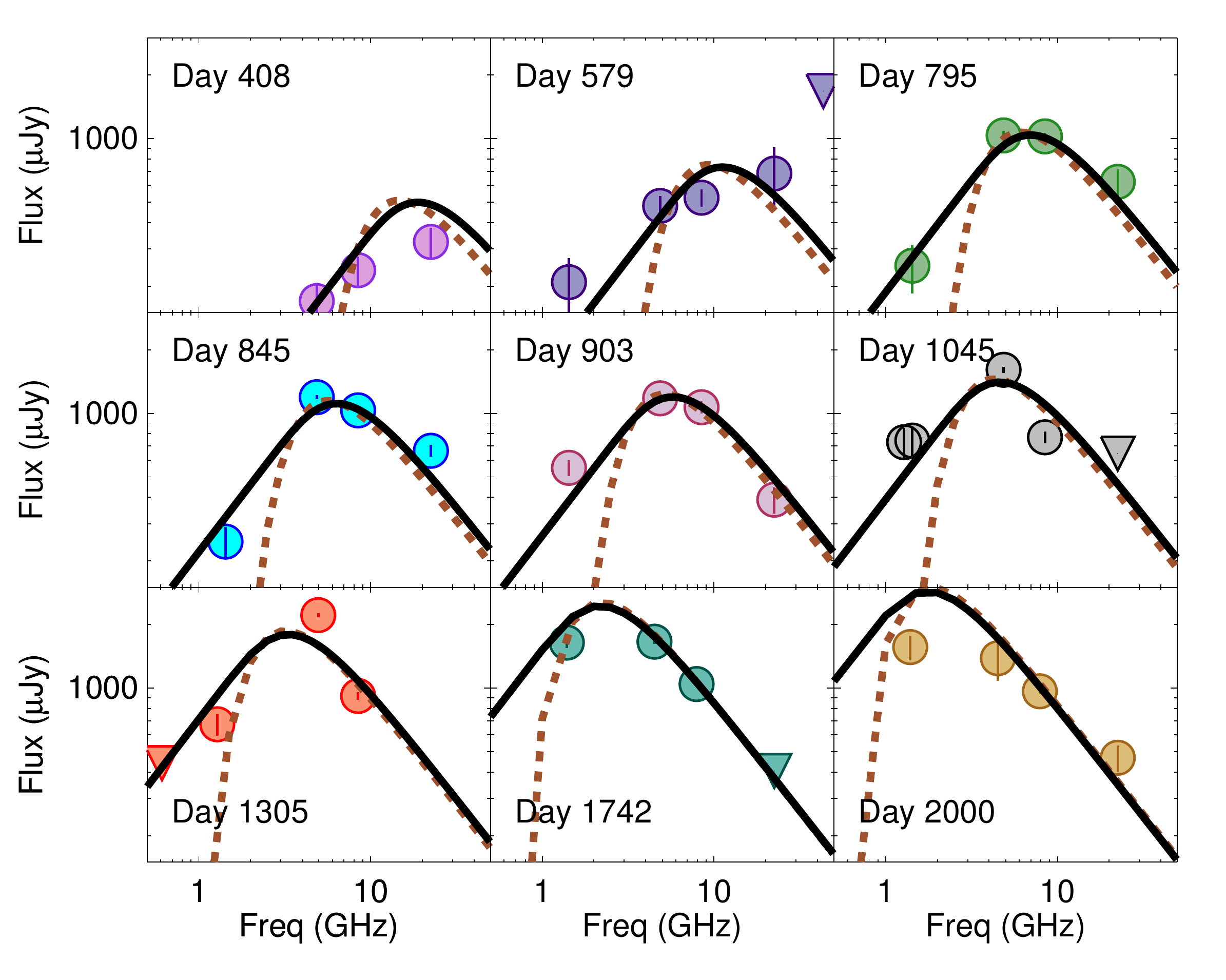} 
 \caption{Spectra of SN 2006jd reproduced from \cite{chandra+12b}. Here internal free-free absorption (black solid lines) best fit the observed absorption.}
   \label{fig4}
\end{center}
\end{figure}

In classic core collapse SNe, the  radio emission is mainly
absorbed either by  the ionized CS medium 
(FFA) or by the in situ relativistic electrons  
(SSA). Observations when the SN is rising in the light curve evolution can disentangle the underlying  absorption mechanism.
The role of  absorption
mechanisms is different for different SNe  \citep{cf03}.  
SNe IIn are expected to have external FFA due to their high density.  
However, in SNe IIn 
due to high density, a radiative cooling shell is formed in the shocked region. The mixing of the cool gas in the forward shocked shell 
may cause some SNe IIn to undergo internal FFA. This indeed has been seen in
SN 2006jd \citep[Fig. \ref{fig4} and ][]{chandra+12b}, SN 1986J \citep{chandra+12b} and SN 1988Z \citep{vandyk+93, williams+02}. 

In SN 2006jd, \citet{chandra+12b}  estimated the mass of the mixed absorbing gas by assuming 
 that the absorbing low temperature ($10^4-10^5$\,K) gas is
in pressure equilibrium with the X-ray emitting gas. They 
showed that  modest amount of $\sim 10^4$ K cool gas, i.e.  $\rm Mass \sim 10^{-8}\rm M_\odot$  could explain internal absorption in SN 2006jd (Fig. \ref{fig4}).

Irrespective of the fact whether absorption is internal or external, it implies very high mass loss rates. 
However, the mass loss rate constraints will become less severe by a factor of $3$ if the wind is clumpy \citep{puls+08}. In addition, 
calculations of mass loss rate are usually done assuming solar metallicity. However, 
metallicity will lower mass loss rate as $\dot M \sim Z^{0.69}$ \citep{vink+01}.
 Interestingly, recent observational findings demonstrated that erratic mass-loss behavior preceding core-collapse extends to H-poor progenitors as well.

\section{Episodic mass loss?}

\begin{figure}
\begin{center}
 \includegraphics[width=0.49\textwidth]{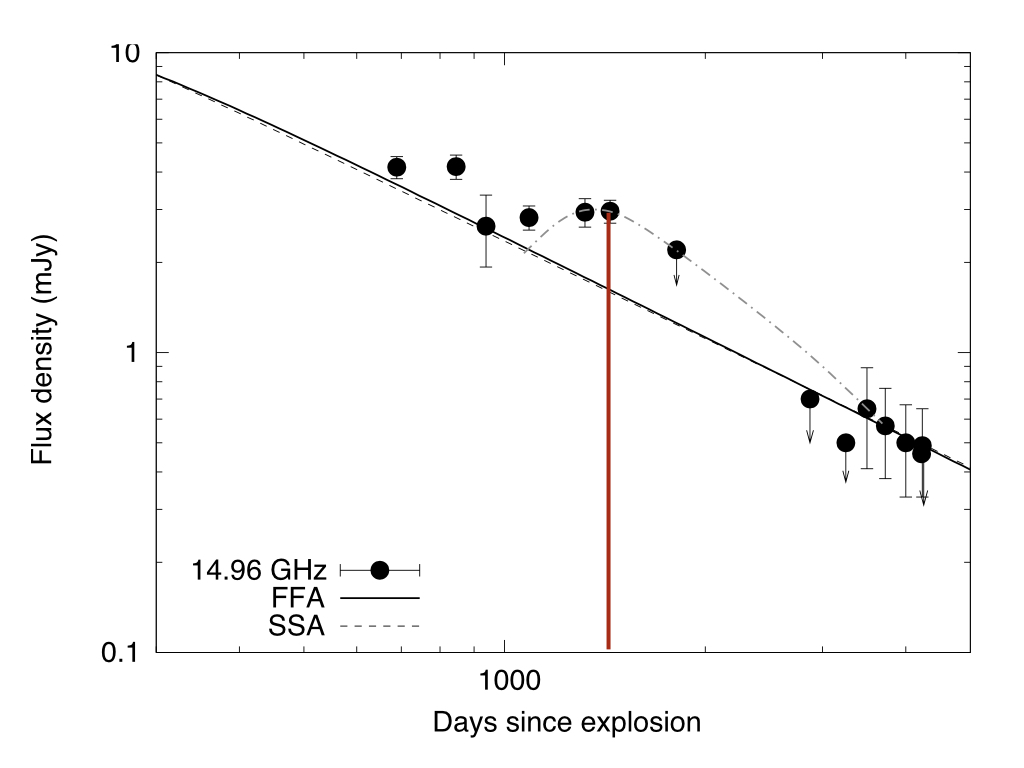} 
  \includegraphics[width=0.49\textwidth]{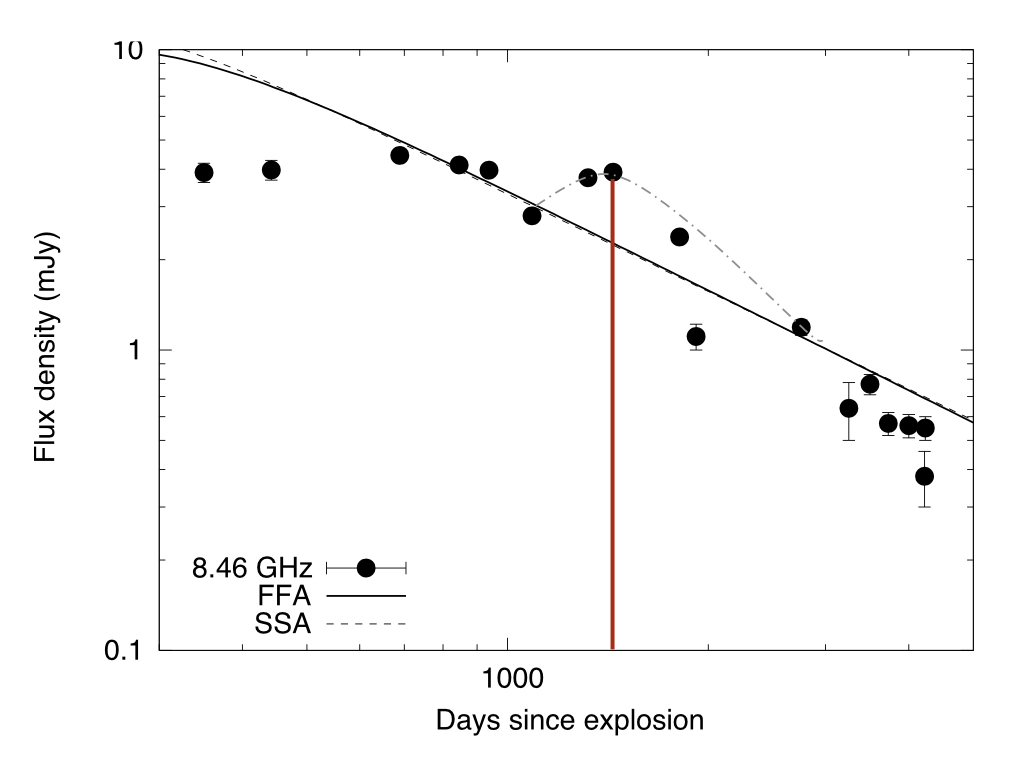}
   \includegraphics[width=0.49\textwidth]{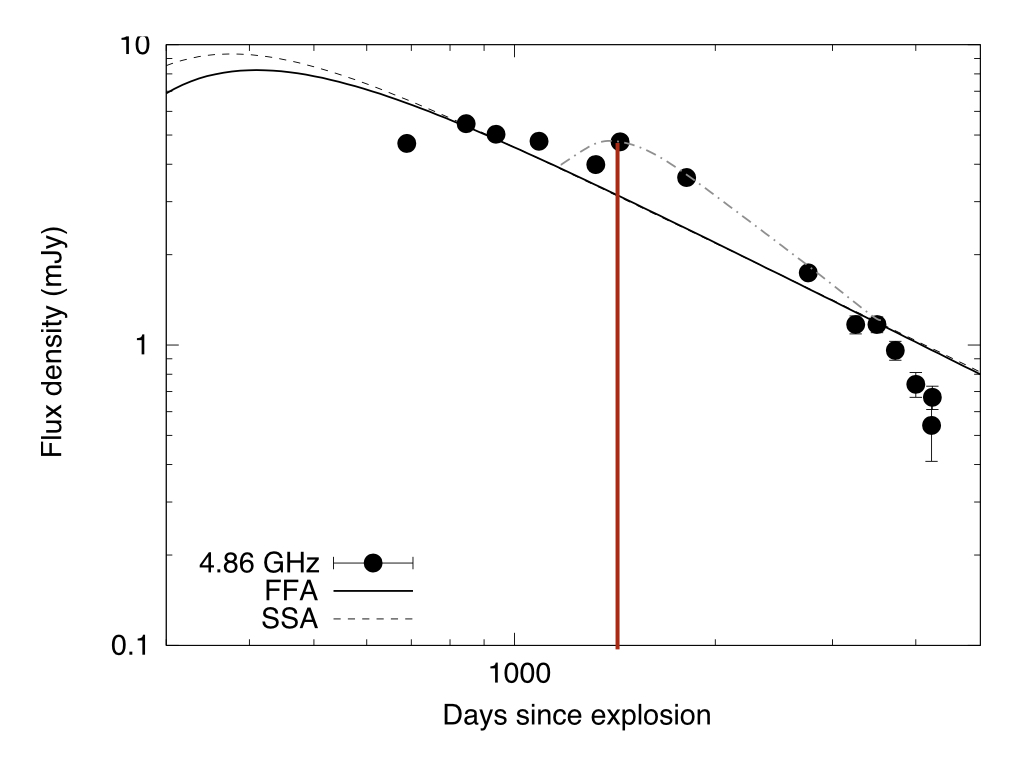}
    \includegraphics[width=0.49\textwidth]{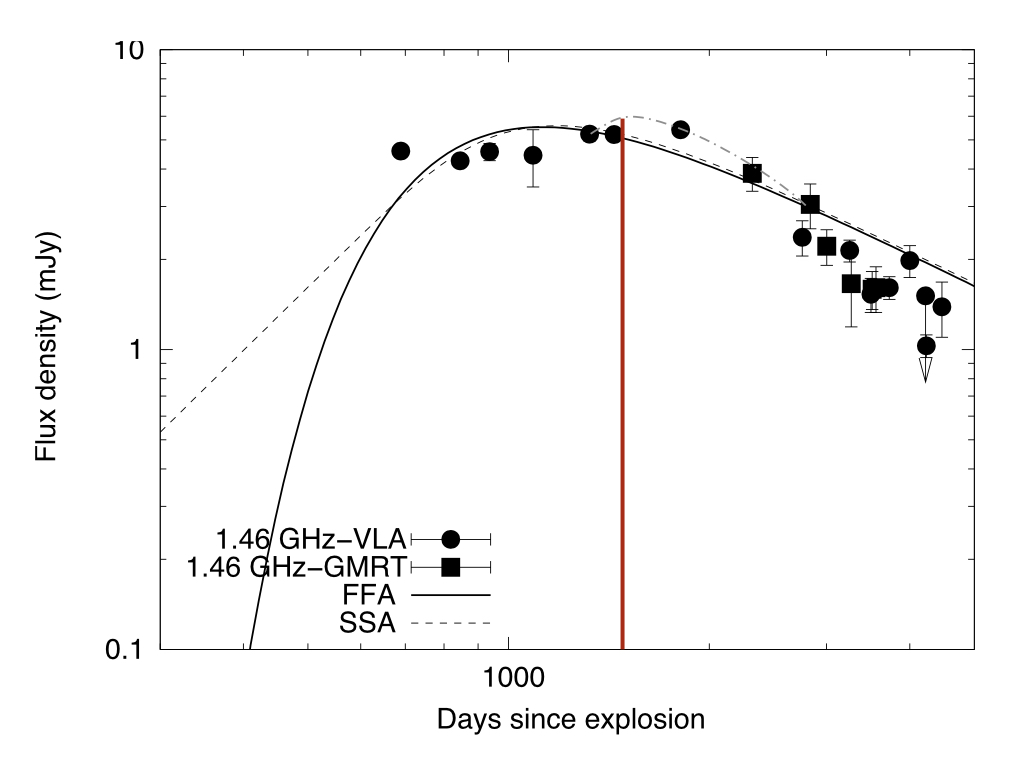}
     \includegraphics[ width=0.49\textwidth]{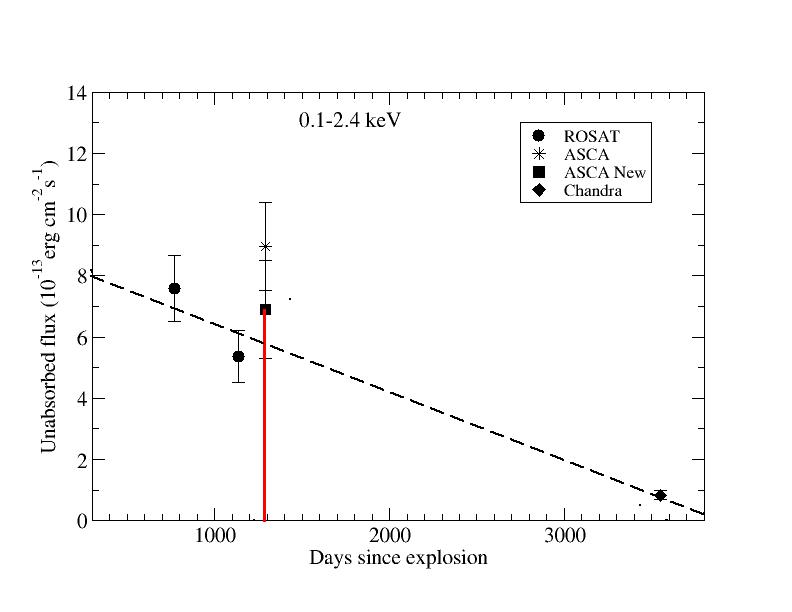}
      \includegraphics[width=0.49\textwidth]{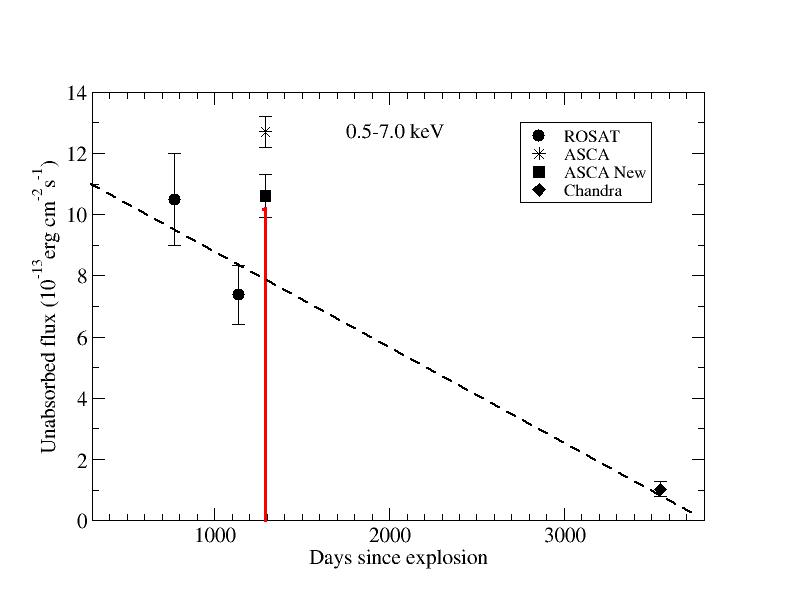}
 \caption{Radio and X-ray light curves of SN 1995N.  A hint of enhanced emission bump in radio and X-ray bands of a Type IIn SN 1995N around $\sim 1300$ days. The figure is reproduced from \citet{chandra+05, chandra+09}.}
   \label{fig6}
\end{center}
\end{figure}

Multiwaeband observations of SNe IIn have shown that the mass loss rate from some SNe IIn may not be steady. Unfortunately it is difficult to disentangle this effect because SNe IIn  are likely to remain 
highly absorbed for a long duration. By the time they become optically thin in radio bands, the strength of synchrotron emission is too  weak to see any modulations due to
changing mass loss.
In addition, one needs a long term follow up to see the episodic mass loss. In X-ray observations, due to the lack of well sampled data, such trends are easy to 
miss. Yet, it seems that some SNe IIn seem to suggest episodic mass loss rate.

In Fig. \ref{fig6}, X-ray and radio light curves are plotted  for a Type IIn SN 1995N \citep{chandra+09}. 
The X-ray bands were chosen to mimic the {\textit ROSAT} and
{\textit ASCA} bands \citep{chandra+05}. The radio data is at the four representative VLA frequencies. Around day $\sim 1300$ onwards,  bumps in the X-ray and radio light curves are seen. 
The long gap in the X-ray light curve does not show the evolution of this bump, however, in the radio data one can clearly see this effect.  Due to the sensitivity limitation of the
telescopes and lack of long term follow up, its not possible to explore the further bumps, if any.

Hints of  bumps near simultaneously in both X-ray and radio bands light curves can be possibly seen in SN 2006jd as well \citep{chandra+12b}.
The radio emission is too weak in SN 2010jl to note any such bumps. However, X-ray data seems to suggest a steady decline followed by a sudden decline
settling down to steady decline again at $\sim 1000$ days \citep{chandra+15}. These trends could be suggestive of the episodic mass loss rates in some SNe IIn and may support the LBV scenario. However, further study and much longer follow ups are needed.
Episodic mass-loss rates have been seen in many type Ib/Ic/IIb SNe too.

\section{Asymmetry in the explosion}

Although in normal core collapse SNe circumstellar interaction  can be successfully described in terms of spherical models, there are indications of a more
complex situation in SNe exploding in dense environments.  
In optical bands, there have been various pieces of evidence suggesting asymmetry in the explosion via polarization observations.
The optical spectropolarimetry observations with Keck telescope showed  a high degree of linear polarization in SN 1998S, implying significant asphericity for its CSM
environment \citep{leonard+00}.
In  SN 1997eg,  spectropolarimetric observations indicated the presence of a flattened disk-like CSM surrounding  a spherical ejecta, which \citet{hoffman+08} 
interpreted as supporting the LBV progenitor scenario.  
\citet{patat+11}  found significant polarization in SN 2010jl two weeks after the discovery, suggesting similarity with SN 1998S and SN 1997eg. 
Many of the models explaining SNe IIn  require  a bipolar wind or
disk like geometry  (see \citet{smith16a} for a review). 

In some SNe IIn,  interpretation of the radio and X-ray data too seem to suggest asymmetry.
In case of SN 2006jd, it  was difficult to reconcile the  X-ray and radio data with spherical models  \citep{chandra+12b}.  The column density of the matter absorbing
 the X-ray emission was found to be  $\sim 50$ times smaller than what was needed to produce the X-ray luminosity.  In addition, the implied   column density from radio 
 observations were also found to be low. \citet{chandra+12b} attempted to explain it with the 
 clumpiness of the CSM clouds, since the absorption column depends on the CSM wind density as  $\propto \rho_{\rm wind}$ and  the X-ray luminosity as  $\propto \rho_{\rm wind}^2$. However, it lead to an inconsistent shock velocity  derived from the observed X-ray spectrum and this model was ruled out.
 They explained  it in a scenario in which   the CSM  clouds interact primarily with the CDS.  This suggested  that there is a global asymmetry in the CSM gas 
 distribution that allowed a low column density in one direction, unaffected by the  dense interaction  taking place over much of the rest of the solid angle as viewed from the SN. 
This would also explain the low  column density to the radio emission too,  because both radio and X-ray are from the same region. 
The lack of external FFA seen in this SN could also be explained in this model. 

SN 2010jl early optical spectra showed the presence of broad emission lines   \citep{smith+12b,fransson+14,ofek+14b}.  This could be 
explained by  an electron scattering optical depth $>1-3$, i.e., a column density $N_H \ge 3 \times 10^{24}$ cm$^{-2}$. This was comparable with the column density  seen  in the first X-ray observations \citep{chandra+12a}.  However, at later epochs ($t > 70$ day),  the X-rays  column density declined to $N_H < 3 \times 10^{24}$ cm$^{-2}$. These constraints are difficult to reconcile with a spherical model of column density $3 \times 10^{24}$ cm$^{-2}$. This may suggest that the  X-rays escape the interaction region, avoiding the high column density
CSM. This situation can be reconciled  well in a scenario for the CSM having a bipolar geometry.

If LBVs indeed are the progenitor of some SNe IIn, they are most likely asymmetric as known   Galactic LBVs show a complex  
CSM structure. In addition, binary evolution can also  lead to aspherical structure \citep[e.g.,][]{mccrayfransson16}.

\section{Summary and open problems}

 In this review, I  have discussed the observational aspects of the ejecta-CSM interaction  mainly  in SNe exploding in dense environments.  The review mainly concentrates on the radio and X-ray emission, with the aim of understanding the nature of intriguing progenitor system of SNe IIn.  
 
 Observations have indicated that in SNe IIn show high X-ray luminosities dominated by high temperatures for years, implying the forward shock is responsible for the 
 dominant X-ray emission. Long term observations of X-ray emission have indicated evolving CSM in some SNe IIn, directly confirming the evidence that
 medium surrounding stars is  modified via the  progenitor star, the properties of which can be studied by the detailed X-ray light curves and spectra. However,  one need to make observations in wide X-ray bands, to avoid biases coming from the narrow X-ray range, and so to  disentangle the
 observational effects versus the ones  coming due to the nature of progenitor wind.
 
 In radio bands, SNe IIn are mostly late emitters.  Radio detected SNe IIn do not tend to be necessarily bright, most
 likely due to absorption.  The absorption mechanism could have a dominant contribution from the cool gas mixed in the radio  emitting region.
This could also be the reason that only 10\% SNe IIn  show radio emission. More systematic studies are needed to remove observational biases and pin point the physical  causes of the  low detection statistics. The property of late time turn on may be causing us to miss  many of them due to lack of sufficiently  observations. With the wealth of new and refurbished sensitive telescopes, a carefully planned sensitive study is needed. 
 A good strategy is  to observe SNe IIn at high frequencies for at least a year before discarding it as a radio non-emitter.

The nature of progenitors of SNe IIn remains the biggest challenge. The mass loss rates needed to explain SNe IIn are extremely high.   Some SNe IIn have indicated episodic mass loss rates. There are hints of eruptive events towards the end stages of evolution.  These effects need to be included in the 
current generations of stellar evolution models. 
 However, it is important to note  here that the mass loss rate constraints will become less severe by a factor of $3$ if the wind is clumpy.  
 
 Many of the massive stars are expected to live in binary systems. Their influence in the stellar evolution need to be explored. 
In addition,  observations have indicated  low progenitor masses in some SNe IIn.  This   has challenged the idea  of interacting SNe arising
from  very massive progenitor stars, and has put a question mark on the LBV progenitors. 
In fact, the very nature of the LBV has been questioned and much work is needed in this direction. A very sensitive observations of the Galactic OB regions and LBV stars 
could throw light on this.

\begin{acknowledgements}
I thank referee for useful comments, which helped improve this review. I acknowledge the support from the Department of Science and Technology via SwaranaJayanti Fellowship award (file no. DST/SJF/PSA-01/2014-15).
The chapter has benefitted immensely from discussions with Roger Chevalier, Claes Fransson, and Nikolai Chugai at various stages.  I am highly indebted to Roger Chevalier for 
proof reading the manuscript.

 \end{acknowledgements}

\bibliographystyle{spbasic}
\bibliography{sn}




\end{document}